\documentclass[pdflatex,prd,onecolumn,showpacs,superscriptaddress,nofootinbib]{revtex4-2}
\usepackage{amssymb,amsmath,color,graphicx,multirow,tabularx,physics}
\usepackage[hyperfootnotes=false]{hyperref}
\hypersetup{
    colorlinks=true,
    linkcolor=blue,
    citecolor=blue,
    urlcolor=blue,
    linktoc=page
}
\usepackage{microtype}

\DeclareRobustCommand{\MBS}{M\"{o}bius Domain Wall Fermions}

\begin{document}
\raggedbottom

\title{Quark Number Susceptibilities and Conserved Charge Fluctuations in $(2+1)$-flavor QCD with {\MBS}}

\author{Jishnu Goswami}
\affiliation{Fakult\"at f\"ur Physik, Universit\"at Bielefeld, 33615 Bielefeld, Germany}

\author{Yasumichi Aoki}
\affiliation{
 RIKEN Center for Computational Science (R-CCS),
 7-1-26, Minatojima Minamimachi, 
Kobe 650-0047, Japan}

\author{Hidenori Fukaya}
\affiliation{Department of Physics, Osaka University, Toyonaka, Osaka 560-0043, Japan}

\author{Shoji Hashimoto}
\affiliation{KEK Theory Center, High Energy Accelerator Research Organization (KEK),
Tsukuba 305-0801, Japan}
\affiliation{School of High Energy Accelerator Science, Graduate University for Advanced Studies (SOKENDAI),
Tsukuba 305-0801, Japan}

\author{Issaku Kanamori}
\affiliation{
 RIKEN Center for Computational Science (R-CCS),
 7-1-26, Minatojima Minamimachi, 
Kobe 650-0047, Japan}

\author{Takashi Kaneko}
\affiliation{KEK Theory Center, High Energy Accelerator Research Organization (KEK),
Tsukuba 305-0801, Japan}
\affiliation{School of High Energy Accelerator Science, Graduate University for Advanced Studies (SOKENDAI),
Tsukuba 305-0801, Japan}

\author{Yoshifumi Nakamura}
\affiliation{
 RIKEN Center for Computational Science (R-CCS),
 7-1-26, Minatojima Minamimachi, 
Kobe 650-0047, Japan}
\author{David Ward}
\affiliation{Department of Physics, Osaka University, Toyonaka, Osaka 560-0043, Japan}
\author{Yu Zhang}
\affiliation{Fakult\"at f\"ur Physik, Universit\"at Bielefeld, 33615 Bielefeld, Germany}

\collaboration{JLQCD Collaboration}
\begin{abstract}
We calculate second- and selected fourth-order conserved-charge fluctuations in $(2+1)$-flavor QCD using M\"obius domain-wall fermions (MDWF) along a line of constant physics. Gauge ensembles were generated for two light-to-strange quark-mass ratios, $m_l/m_s=1/10$ and $1/27.4$, corresponding to heavier-than-physical and physical pion masses, respectively. For $m_l/m_s=1/10$, calculations were carried out on lattices with temporal extents $N_\tau=12$ and $16$, enabling an assessment of lattice-spacing effects at heavier pion mass. For $m_l/m_s=1/27.4$, calculations were performed at $N_\tau=12$, allowing us to study the light-quark-mass dependence down to the physical point. {Several second-order conserved-charge fluctuations show qualitative consistency with the QMHRG2020 hadron resonance gas baseline below the pseudocritical region within present uncertainties.} Across the crossover region, these observables rise rapidly and tend toward their Stefan--Boltzmann limits. Selected fourth-order cumulants were also computed at the physical pion mass. Although these observables are statistically more demanding, several channels with controlled uncertainties permit a first comparison with hadron resonance gas calculations.
\end{abstract}

\maketitle
\begin{flushleft}
  \footnotesize Report~No.\ KEK-CP-0409 \\
   \footnotesize Report~No.\ OU-HET-1279
\end{flushleft}
\section{Introduction}\label{sec:intro}

Understanding the phase structure of Quantum Chromodynamics (QCD) at finite temperature and baryon density remains a central problem in strong-interaction physics. At vanishing baryon chemical potential, lattice QCD calculations have established that the transition from the low-temperature hadronic regime to the high-temperature quark-gluon plasma is a smooth crossover rather than a genuine phase transition \cite{Aoki:2006we}. At larger baryon densities, it has long been conjectured that this crossover may turn into a first-order phase transition, implying the existence of a critical endpoint in the QCD phase diagram \cite{Stephanov:2004wx}. Event-by-event fluctuations of conserved charges measured in heavy-ion collision experiments at RHIC and the LHC provide an important experimental probe of this phase structure.

On the theory side, lattice QCD provides first-principles results for fluctuations of baryon number, electric charge, and strangeness through derivatives of the pressure with respect to the corresponding chemical potentials. Because direct simulations at nonzero baryon chemical potential are obstructed by the sign problem, such fluctuations are commonly studied using indirect approaches, most notably Taylor expansions around vanishing chemical potential or analytic continuation from imaginary chemical potentials. In this work, we employ the Taylor-expansion method to compute conserved-charge fluctuations in finite-temperature QCD. These observables provide equilibrium-QCD input relevant for the interpretation of heavy-ion data and for assessing the hadron resonance gas (HRG) baseline in the crossover region.

Lattice calculations of conserved-charge fluctuations have so far been carried out predominantly with staggered fermions. In particular, continuum-extrapolated results for second-order fluctuations are available from simulations with highly improved staggered quarks (HISQ) and stout link actions using lattices with $N_\tau=6$, 8, 12, and 16~\cite{Bellwied:2019pxh,Bollweg:2021vqf}. {At finite lattice spacing, taste-symmetry violations in staggered formulations modify the light pseudoscalar spectrum. Such effects can be particularly relevant for pion-sensitive observables, including electric-charge fluctuations~\cite{HotQCD:2018pds}. Calculations with an alternative fermion discretization therefore provide a useful check of finite-cutoff systematics.}

In this work, we study conserved-charge fluctuations using M\"obius domain-wall fermions (MDWF)~\cite{Brower:2012vk,Kaneko:2013jla,Aoki:2021kbh,Aoki:2023fmn}, which offer improved control of chiral symmetry at finite lattice spacing. Although MDWF simulations are computationally more demanding than staggered-fermion calculations, recent algorithmic and computational advances have made realistic thermodynamic studies feasible. We analyze gauge ensembles generated along a line of constant physics for two light-to-strange quark-mass ratios, $m_l/m_s=1/10$ and $1/27.4$, corresponding to heavier-than-physical and physical pion masses, respectively. This allows us to investigate both lattice-spacing effects at heavier pion mass and the light-quark-mass dependence of conserved-charge fluctuations down to the physical point. {Our main focus is on second-order fluctuations, including comparisons among light-, strange-, and baryon-dominated channels.} We also present results for selected fourth-order cumulants at the physical pion mass.

The line of constant physics and finite-temperature ensembles used in this work were established in Refs.~\cite{Aoki:2021kbh,Aoki:2023fmn}, where chiral observables were analyzed with the same action setup. Preliminary results for charge fluctuations were reported in Ref.~\cite{Goswami:2025euh}. In Secs.~\ref{sec:lattset} and \ref{sec:qnsmeas}, we describe the lattice setup, quark-mass definitions, and the methods used to compute quark number susceptibilities and conserved-charge fluctuations. Results for second-order fluctuations at vanishing chemical potential are presented in Secs.~\ref{sec:qnsresults} and \ref{sec:cnfresults}, while selected fourth-order cumulants and comparisons with QMHRG2020 are discussed in Sec.~\ref{sec:cnfresultsfourthorder}. Unless stated otherwise, temperatures are quoted in MeV and susceptibilities are given in standard dimensionless normalization.
\section{Lattice setup}
\label{sec:lattset}
Gauge ensembles for $(2+1)$-flavor QCD were generated using M\"obius domain-wall fermions (MDWF) and the tree-level Symanzik-improved gauge action with the rational hybrid Monte Carlo (RHMC) algorithm~\cite{Aoki:2021kbh,Aoki:2023fmn}. The MDWF kernel is
\begin{equation}
    D_{\mathrm{M}} = \frac{(b+c)D_{\mathrm{Wilson}}(-M_5)}{2 + (b-c)D_{\mathrm{Wilson}}(-M_5)},
\end{equation}
with parameters $b = 3/2$, $c = 1/2$, $M_5 = 1.0$, and fifth-dimensional extent $L_s = 12$. Here, $D_{\mathrm{Wilson}}$ denotes the four-dimensional Wilson Dirac operator. To reduce residual chiral-symmetry breaking, three levels of stout-link smearing are applied~\cite{Hashimoto:2014gta}. The lattice spacing $a(\beta)$ is set through the Wilson-flow scale $t_0$~\cite{Aoki:2021kbh}.

Throughout this section, $m_f$ ($f\in\{l,s\}$) denotes the input bare quark mass used in the simulation. Because $L_s$ is finite, the quark mass receives an additive residual-mass correction,
\begin{equation}
m_f^{\mathrm{latt}} = m_f + m_{\mathrm{res}}(\beta),
\end{equation}
where $m_f^{\mathrm{latt}}$ is the effective lattice quark mass entering the definition of the line of constant physics (LCP).

\begin{table}[tbp]
\centering
\setlength{\tabcolsep}{8pt}
\begin{tabular}{ccc@{\hspace{3em}}ccc}
\hline\hline
$\beta$ & $m_s^{\mathrm{latt}}$ & $m_{\mathrm{res}}$ & $\beta$ & $m_s^{\mathrm{latt}}$ & $m_{\mathrm{res}}$ \\
\hline
4.02 & 0.0704776 & 0.005913 & 4.17 & 0.0388152 & 0.000415 \\
4.04 & 0.0639011 & 0.003989 & 4.19 & 0.0364795 & 0.000297 \\
4.06 & 0.0583505 & 0.002741 & 4.21 & 0.0343748 & 0.000212 \\
4.08 & 0.0536129 & 0.001903 & 4.23 & 0.0324666 & 0.000152 \\
4.10 & 0.0495275 & 0.001322 & 4.25 & 0.0307270 & 0.000109 \\
4.13 & 0.0443603 & 0.000814 & 4.27 & 0.0291331 & 0.000070 \\
4.15 & 0.0414244 & 0.000581 & 4.30 & 0.0269749 & 0.000047 \\
\hline\hline
\end{tabular}
\caption{$m_s^{\mathrm{latt}}$ and $m_{\mathrm{res}}$ over the full $\beta$ range used in this work. {
The residual-mass determination follows the zero-temperature MDWF analysis of Ref.~\cite{Colquhoun:2022atw}; its use in the finite-temperature analysis is discussed in Ref.~\cite{Aoki:2023fmn}.}}
\label{tab:mres}
\end{table}

\begin{table*}[htbp]
  \centering
  \begin{ruledtabular}
  \begin{tabular}{cccc|cccc|cccc}
    \multicolumn{4}{c|}{$m_l/m_s=1/10;\;24^3\times12$} 
    & \multicolumn{4}{c|}{$m_l/m_s=1/10;\;32^3\times16$} 
    & \multicolumn{4}{c}{$m_l/m_s=1/27.4;\;36^3\times12$} \\
    $\beta$ & $T$ [MeV] & \#Configs & Separation
    & $\beta$ & $T$ [MeV] & \#Configs & Separation
    & $\beta$ & $T$ [MeV] & \#Configs & Separation \\
    \hline
    4.02 & 130.30 & 194 & 100 & 4.10 & 127.05 & 187 & 100 & 4.04 & 139.96 & 195 & 100 \\
    4.04 & 139.96 & 198 & 100 & 4.13 & 138.23 & 190 & 100 & 4.06 & 149.70 & 187 & 100 \\
    4.06 & 149.70 & 198 & 100 & 4.15 & 145.75 & 186 & 100 & 4.07 & 154.60 & 174 & 100 \\
    4.08 & 159.52 & 196 & 100 & 4.17 & 153.31 & 196 & 100 & 4.08 & 159.52 & 194 & 100 \\
    4.10 & 169.39 & 181 & 100 & 4.19 & 160.94 & 189 & 100 & 4.11 & 174.35 & 202 & 100 \\
    4.11 & 174.35 & 186 & 100 & 4.21 & 168.64 & 185 & 100 & 4.17 & 204.42 & 204 & 100 \\
    4.13 & 184.31 & 187 & 100 & 4.23 & 176.43 & 188 & 100 & -- & -- & -- & -- \\
    4.15 & 194.33 & 188 & 100 & 4.25 & 184.31 & 184 & 100 & -- & -- & -- & -- \\
    4.17 & 204.42 & 189 & 100 & 4.27 & 192.30 & 193 & 100 & -- & -- & -- & -- \\
    -- & -- & -- & -- & 4.30 & 204.52 & 198 & 100 & -- & -- & -- & -- \\
  \end{tabular}
  \end{ruledtabular}
  \caption{Parameters of the finite-temperature ensembles used in this work. Measurements were performed on configurations separated by 100 molecular-dynamics trajectories. {The mass ratios in the column headings denote the target effective lattice ratios, $m_l^{\mathrm{latt}}/m_s^{\mathrm{latt}}$.}}
  \label{tab:ensembles}
\end{table*}

The simulations were performed along a line of constant physics defined by
\begin{equation}
\frac{m_l^{\mathrm{latt}}}{m_s^{\mathrm{latt}}}
=
\frac{m_l + m_{\mathrm{res}}(\beta)}{m_s + m_{\mathrm{res}}(\beta)}
= \mathrm{const}.
\end{equation}
We define the line of constant physics by keeping the renormalized strange-quark mass constant in physical units, with
\[
m_s^{\mathrm{phys}} = Z_m(\beta)\, m_s^{\mathrm{latt}}\, a^{-1}(\beta),
\]
where $m_s^{\mathrm{phys}} = 92~\mathrm{MeV}$~\cite{FlavourLatticeAveragingGroupFLAG:2021npn} and $\beta$ is the gauge coupling. The determinations of $a(\beta)$ and $Z_m(\beta)$ are given in \cite{Aoki:2021kbh}.
We consider two target mass ratios, $m_l^{\mathrm{latt}}/m_s^{\mathrm{latt}}=1/10$ and $1/27.4$, corresponding to heavier-than-physical and physical pion masses, respectively. For brevity, we refer to these ensembles as $m_l/m_s=1/10$ and $m_l/m_s=1/27.4$ in the figures, tables, and discussion below.

\begin{figure}[t]
    \centering
    \includegraphics[width=0.38\textwidth]{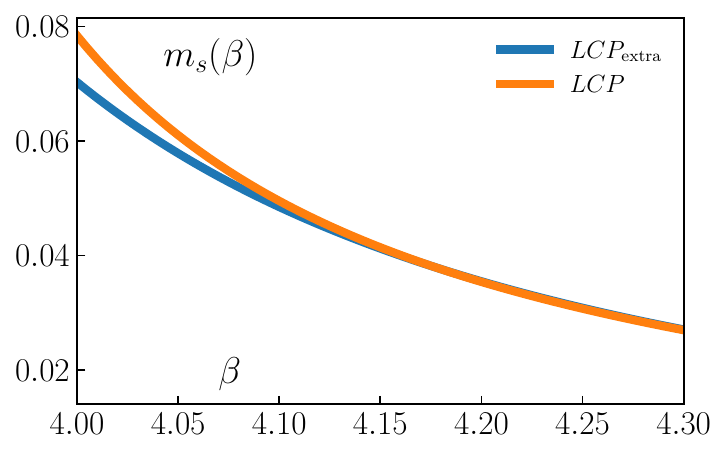}
    \includegraphics[width=0.38\textwidth]{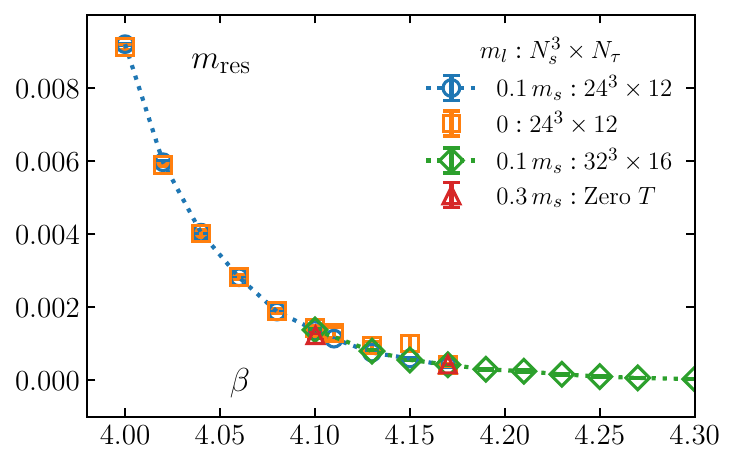}
    \caption{Left: Strange-quark mass $m_s(\beta)$ as a function of the gauge coupling. On LCP$_{\rm extrap}$, the kaon mass deviates by about $10\%$ at $\beta=4.00$, whereas on the corrected LCP the deviation is below $1\%$. Statistical errors are smaller than the line width. Right: Residual mass $m_{\mathrm{res}}$ as a function of $\beta$. The residual mass is nearly independent of quark mass and volume. Zero-temperature data are taken from Ref.~\cite{Colquhoun:2022atw}.}
    \label{fig:mres}
\end{figure}

At $\beta=4.00$, where zero-temperature reference data were not yet available, the strange-quark mass was extrapolated from higher-$\beta$ data ($\beta \ge 4.10$). This led to a mistuned line of constant physics, denoted by LCP$_{\rm extrap}$, for which the strange-quark mass was underestimated by about $10\%$. A corrected LCP was established later using dedicated zero-temperature simulations
{~\cite{Aoki:2023fmn,Goswami:2025euh}}.  Figure~\ref{fig:mres} (left) compares the extrapolated and corrected LCPs, while Fig.~\ref{fig:mres} (right) shows the residual mass $m_{\mathrm{res}}(\beta)$, which is nearly independent of the input quark mass.

Simulations were carried out with temporal extents $N_\tau=12$ for $4.02 \le \beta \le 4.17$ and $N_\tau=16$ for $4.10 \le \beta \le 4.30$, corresponding to temperatures in the range $127 \lesssim T \lesssim 205$~MeV. The ensemble-generation strategy is summarized as follows:

\begin{itemize}
\item \textbf{Heavier pion mass ($m_l/m_s=1/10$):}
Initial ensembles at both $N_\tau=12$ and $16$ were generated on LCP$_{\rm extrap}$, after which $m_{\mathrm{res}}(\beta)$ was measured. For $N_\tau=12$, new ensembles were generated with input bare masses obtained by subtracting $m_{\mathrm{res}}(\beta)$ from the target lattice masses on LCP$_{\rm extrap}$, and mass reweighting was then used to shift observables to the corrected LCP. For $N_\tau=16$, where $m_{\mathrm{res}}$ is smaller, mass reweighting was applied directly to the original ensembles to account for both the residual-mass correction and the LCP mistuning. All results quoted for $m_l/m_s=1/10$ correspond to the corrected LCP.

\item \textbf{Physical pion mass ($m_l/m_s=1/27.4$):}
These ensembles were generated directly on the corrected LCP. The input bare masses were chosen by subtracting $m_{\mathrm{res}}(\beta)$ from the target lattice masses on the corrected LCP. The residual masses determined in the $m_l/m_s=1/10$ analysis were used here as well, since $m_{\mathrm{res}}$ is nearly independent of the light-quark mass. No additional mass reweighting was required.
\end{itemize}
For the $m_l/m_s=1/10$, $N_\tau=16$ ensembles at $\beta=4.27$ and $4.30$, the input bare masses were not additionally corrected for $m_{\rm res}$. The resulting mistuning is small, about $2\%$ for the light-quark mass and $0.2\%$ for the strange-quark mass.

Table~\ref{tab:mres} summarizes $m_s^{\mathrm{latt}}$ and $m_{\mathrm{res}}$ over the full $\beta$ range, and Table~\ref{tab:ensembles} lists the finite-temperature ensembles analyzed in this work. Statistical uncertainties are estimated using jackknife resampling. Details of the measurement of Taylor-expansion coefficients of the pressure up to fourth order are given in Sec.~\ref{sec:qnsmeas}.
\section{Quark number susceptibilities up to fourth order}
\label{sec:qnsmeas}

In this section, we summarize the Taylor-expansion formalism used to compute quark number susceptibilities up to fourth order. These quantities provide the basic building blocks for the conserved-charge fluctuations discussed in the following sections.

For $(2+1)$-flavor QCD with two light quarks $(u,d)$ and one strange quark $(s)$, the pressure can be expanded in the quark chemical potentials as
\begin{equation}
\frac{P}{T^4}
=
\sum_{i+j+k=\mathrm{even}}
\frac{\chi_{ijk}^{uds}}{i!\,j!\,k!}\,
\hat\mu_u^i\,\hat\mu_d^j\,\hat\mu_s^k,
\qquad
\chi_{ijk}^{uds}
=
\frac{1}{VT^3}
\frac{\partial^{\,i+j+k}\ln Z}
{\partial\hat\mu_u^i\,\partial\hat\mu_d^j\,\partial\hat\mu_s^k}
\Bigg|_{\vec{\hat\mu}=0},
\label{chiTaylor}
\end{equation}
where $\hat\mu_f \equiv \mu_f/T$ for $f\in\{u,d,s\}$. The coefficients $\chi_{ijk}^{uds}$ are the quark number susceptibilities evaluated at vanishing chemical potentials.

For the evaluation of these derivatives on the lattice, we introduce the
dimensionless lattice chemical potential
\[
\bar\mu_f \equiv a\mu_f = \frac{\hat\mu_f}{N_\tau},
\]
where $T^{-1}=aN_\tau$.
The quark chemical potential is introduced in the MDWF action through the
standard temporal-link prescription
\[
(1 \pm \gamma_4) U_{\pm 4}(x)
\;\to\;
(1 \pm \gamma_4)e^{\pm \bar\mu_f}\,U_{\pm 4}(x).
\]
\cite{Bloch:2007xi,Brower:2012vk}.
The partition function can then be written as
\begin{align}
	Z
	&=
	\int DU \;
	\prod_{f=u,d,s}\det \mathcal{M}(m_f,\bar\mu_f)\;
	e^{-S_g},
	\nonumber\\
	\det \mathcal{M}(m_f,\bar\mu_f)
	&=
	\frac{\det D(m_f,\bar\mu_f)}
	{\det D(m_{\rm PV},\bar\mu_f)}.
	\label{eq:partitionfunc}
\end{align}

where $m_{\rm PV}=1$ is used for the Pauli--Villars subtraction.

We express the susceptibilities in terms of
\begin{equation}
	D_n^f
	\equiv
	\left.
	\frac{\partial^n}{\partial \bar\mu_f^n}
	\ln \det \mathcal{M}(m_f,\bar\mu_f)
	\right|_{\vec{\bar\mu}=0},
	\qquad
	f\in\{u,d,s\}.
	\label{eq:Dn}
\end{equation}

Since $\hat\mu_f=N_\tau\bar\mu_f$, converting the lattice derivatives
with respect to $\bar\mu_f$ to derivatives with respect to $\hat\mu_f$
gives the factors of $N_\tau$ appearing below.

Throughout this paper, $D(m_f)$ denotes the MDWF Dirac operator with quark mass $m_f$ at vanishing chemical potential unless stated otherwise. We use the standard identities
\[
\frac{\partial}{\partial\bar\mu}\ln\det D
=
\Tr\!\left[D^{-1}\frac{\partial D}{\partial\bar\mu}\right],
\]
\[
\frac{\partial^2}{\partial\bar\mu^2}\ln\det D
=
\Tr\!\left[D^{-1}\frac{\partial^2 D}{\partial\bar\mu^2}\right]
-
\Tr\!\left[D^{-1}\frac{\partial D}{\partial\bar\mu}
D^{-1}\frac{\partial D}{\partial\bar\mu}\right],
\]
and analogous expressions for higher derivatives. The complete third- and fourth-derivative formulas are given in Appendix~A.

\subsection{Second- and fourth-order quark number susceptibilities}

The coefficients $\chi_{ijk}^{uds}$ with $i+j+k=2$ correspond to second-order susceptibilities. Using the shorthand
\[
\chi_{200}^{uds}\equiv\chi_2^u,
\qquad
\chi_{110}^{uds}\equiv\chi_{11}^{ud},
\qquad
\chi_{101}^{uds}\equiv\chi_{11}^{us},
\]
the diagonal and off-diagonal second-order susceptibilities are
\begin{equation}
\chi_2^f
=
\frac{N_\tau}{N_\sigma^3}
\left[
\langle D_2^f\rangle
+
\langle (D_1^f)^2\rangle
\right],
\qquad
\chi_{11}^{fg}
=
\frac{N_\tau}{N_\sigma^3}
\langle D_1^f D_1^g\rangle,
\qquad
(f\neq g).
\label{eq:second_order}
\end{equation}
For degenerate light quarks, $m_u=m_d$, the $u$- and $d$-quark Dirac operators are identical on each gauge configuration. Consequently, $D_1^u=D_1^d$ configuration by configuration, and therefore $(D_1^u)^2 = D_1^u D_1^d$.

Fourth-order cumulants are constructed from the trace operators $D_n^f$ as
\begin{align}
\chi_4^f
&=
\frac{1}{N_\tau N_\sigma^3}
\left[
Z_4^f - 3(Z_2^f)^2
\right],
\\
\chi_{31}^{fg}
&=
\frac{1}{N_\tau N_\sigma^3}
\left[
Z_{31}^{fg} - 3 Z_2^f Z_{11}^{fg}
\right],
\qquad (f\neq g),
\\
\chi_{22}^{fg}
&=
\frac{1}{N_\tau N_\sigma^3}
\left[
Z_{22}^{fg} - Z_2^f Z_2^g - 2(Z_{11}^{fg})^2
\right],
\qquad (f\neq g),
\\
\chi_{211}^{fgh}
&=
\frac{1}{N_\tau N_\sigma^3}
\left[
Z_{211}^{fgh} - 2 Z_{11}^{fg} Z_{11}^{fh} - Z_{11}^{gh} Z_2^f
\right],
\qquad (f\neq g\neq h).
\end{align}
The explicit expressions for the $Z$-combinations in terms of the $D_n^f$ are collected in Appendix~A.
\section{Quark number susceptibilities}
\label{sec:qnsresults}

In this section, we discuss the numerical strategy used for the evaluation of quark number susceptibilities and summarize the resulting second- and fourth-order observables. Results are presented for the heavier light-quark mass, $m_l=m_s/10$, and for the physical light-quark mass, $m_l=m_s/27.4$. Throughout the discussion, cutoff effects are characterized in terms of the temporal extent $N_\tau$, with larger $N_\tau$ corresponding to finer lattice spacing at fixed temperature. For $m_l=m_s/10$, calculations were performed on lattices with $N_\tau=12$ and 16 and aspect ratio $N_\sigma/N_\tau=2$. For $m_l=m_s/27.4$, calculations were carried out on lattices with $N_\tau=12$ and aspect ratio $N_\sigma/N_\tau=3$.

The evaluation of quark number susceptibilities requires stochastic estimates of traces involving derivatives of the fermion matrix. In practice, the dominant source of stochastic noise arises from disconnected contributions involving powers of $D_1^f$, in particular $(D_1^f)^n$ with $n=2,4,\ldots$. We therefore use Gaussian random sources together with dilution and unbiased estimators to control the stochastic uncertainty. For example,
\begin{eqnarray}
D_1^f &=& \mathrm{Tr} \left[ D(m_f)^{-1}\frac{dD}{d\bar\mu_f} \right] - \mathrm{Tr} \left[ D(m_{\text{PV}})^{-1}\frac{dD}{d\bar\mu_f} \right] \nonumber\\
&\simeq& \frac{1}{N_n}\sum_{j=1}^{N_n} \left[\eta^{\dagger}_{j} D(m_f)^{-1}\frac{dD}{d\bar\mu_f} \eta_{j} - \eta^{\dagger}_{j} D(m_{\text{PV}})^{-1}\frac{dD}{d\bar\mu_f} \eta_{j}\right],
\label{eq:dfeqnoise}
\end{eqnarray}
where $\eta_j$ denotes a Gaussian random source and $N_n$ is the number of noise vectors. In the estimator for $D_1^f$, we use identical noise vectors for the fermion and Pauli--Villars parts in order to reduce the stochastic variance in $(D_1^f)^2$.

\subsection{Dilution of noise vectors}

To further reduce stochastic noise, we employ dilution, i.e.\ the decomposition of a noise vector into orthogonal subspaces~\cite{Foley:2005ac}. In this case,
\begin{eqnarray}
D_1^f &\simeq& \frac{1}{N_n}\sum_{j=1}^{N_n} \left[\sum_{a=1}^{N}\eta^{\dagger}_{aj} D(m_f)^{-1}\frac{dD(m_f)}{d\bar\mu_f} \eta_{aj} - \sum_{a=1}^{N}\eta^{\dagger}_{aj} D(m_{\text{PV}})^{-1}\frac{dD(m_{\text{PV}})}{d\bar\mu_f} \eta_{aj}\right],
\label{eq:dilution}
\end{eqnarray}
where $\eta_{aj}$ denotes a diluted Gaussian source. We consider three dilution schemes:
\begin{enumerate}
\item \textbf{Even-odd dilution}: decomposition into even and odd lattice sites ($N=2$),
\item \textbf{Spin dilution}: decomposition into the four spin components ($N=4$)~\cite{QCDSF:2011rov},
\item \textbf{Time-slice dilution}: decomposition into four subsets according to $t \bmod 4$~\cite{Foley:2005ac}.
\end{enumerate}

\begin{figure}[!htbp]
    \centering
    \includegraphics[width=0.42\textwidth]{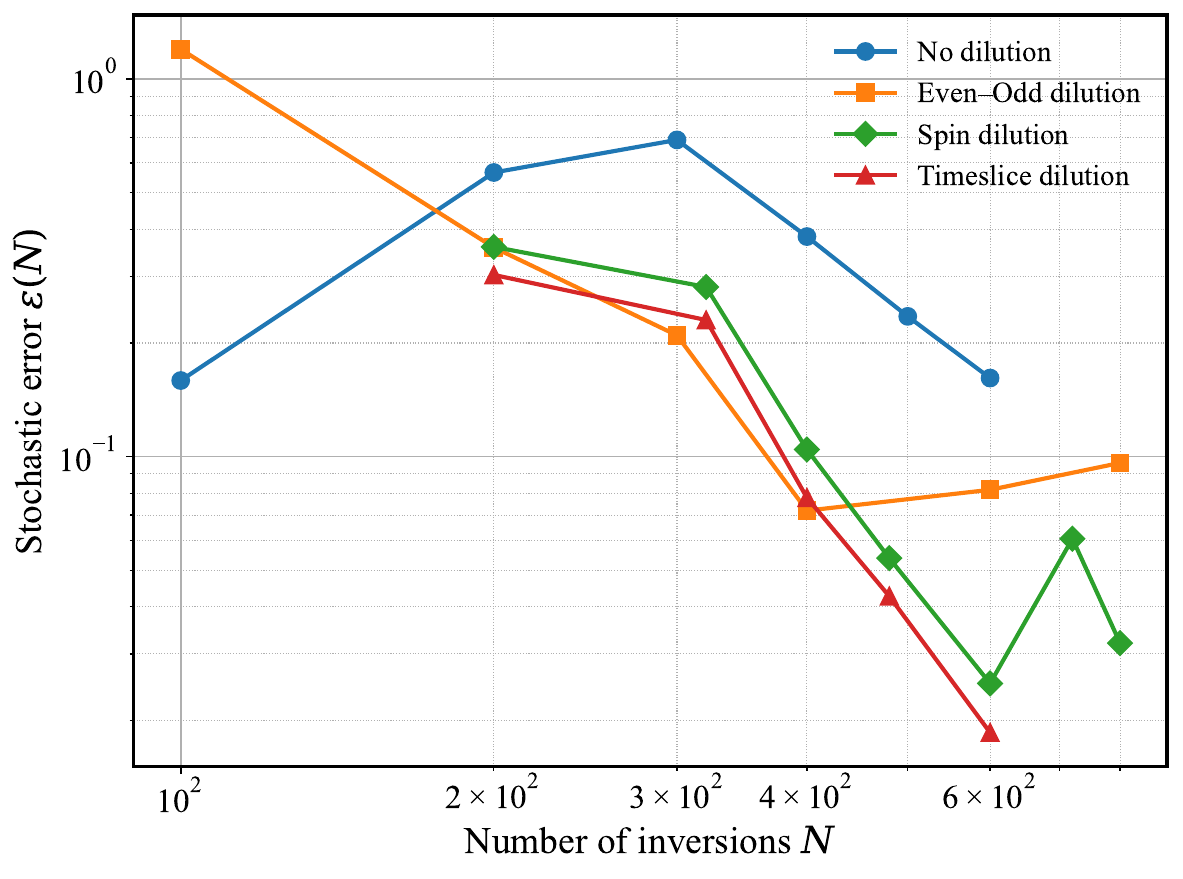}
    \includegraphics[width=0.42\textwidth]{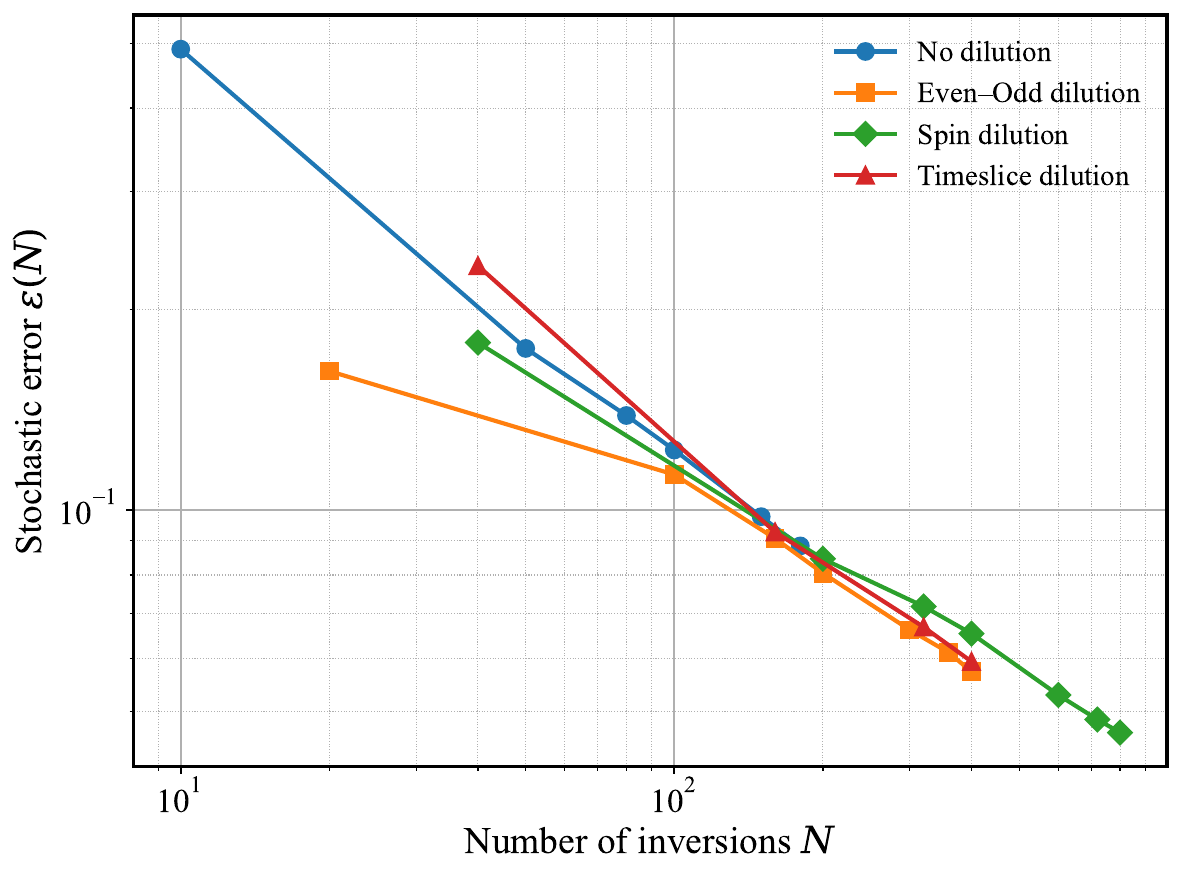}
    \caption{Stochastic errors for a representative gauge configuration as functions of the total number of inversions. Left: error in $(D_1^u)^2$. Right: error in $D_2^u$. {The comparison is made at fixed inversion cost; the errors are obtained by jackknife resampling over independent random-source estimates on the same gauge configuration.}}
    \label{fig:stochniose}
\end{figure}

Spin dilution and time-slice dilution require four times more inversions per noise vector, while even-odd dilution requires a factor of two more inversions. Figure~\ref{fig:stochniose} compares the stochastic error at fixed inversion cost for a representative gauge configuration. {The errors shown in this figure are stochastic errors at fixed gauge field, estimated by jackknife resampling over the independent random-source estimates.} For a given number of inversions, dilution reduces the stochastic uncertainty in $(D_1^u)^2$ by roughly a factor of $2$--$3$, while the gain for $D_2^u$ is more modest. Among the schemes considered here, spin and time-slice dilution are the most cost-effective. In practice, we use spin dilution for the $N_\tau=12$ ensembles and time-slice dilution for the $N_\tau=16$ ensembles.

\begin{figure}[!htbp]
    \centering
    \includegraphics[width=0.58\textwidth]{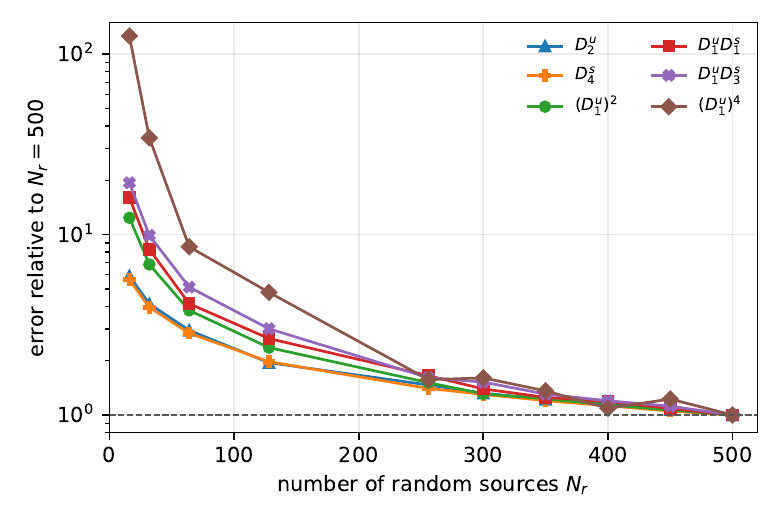}
    \caption{Stochastic-source check for one physical-mass configuration at $\beta=4.06$ ($T=149.7~\mathrm{MeV}$). The vertical axis is the ratio $\sigma(N_r)/\sigma(500)$; values above one indicate larger stochastic error than in the final $N_r=500$ measurement. The range $N_r=300$--500 is sampled more densely to show the behavior near the production choice. The examples include diagonal, off-diagonal, and fourth-order trace structures.}
    \label{fig:noise-vector-dependence}
\end{figure}

As an additional check on the dependence on the number of stochastic sources, Fig.~\ref{fig:noise-vector-dependence} shows the stochastic error at smaller $N_r$ divided by the error obtained with the production choice, $N_r=500$. Thus a value of 5 on the vertical axis means an error five times larger than the final 500-source estimate. The examples include light--strange off-diagonal products, $D_1^uD_1^s$ and $D_1^uD_3^s$, as well as second- and fourth-order trace structures. The error decreases rapidly at small $N_r$, with the strongest effect seen for fourth-order structures containing powers of $D_1^u$, and changes more slowly in the range $N_r=300$--500. We use this figure as a source-number dependence diagnostic. The stability of the final observable errors under blocking is discussed in Appendix~\ref{app:resampling}.

\subsection{Trace measurements}

The traces entering the quark number susceptibilities were measured on the ensembles listed in Table~\ref{tab:ensembles}. Statistical errors were estimated with the jackknife method and include both gauge-field fluctuations and stochastic uncertainties from the finite number of noise vectors. As shown below, disconnected contributions, especially those entering off-diagonal susceptibilities such as $\chi_{11}^{ud}$ and $\chi_{11}^{us}$, dominate the total error budget.

For the heavier light-quark mass, $m_l=m_s/10$, we computed second-order quark number susceptibilities on lattices with $N_\tau=12$ and 16. In both cases, 200 Gaussian random sources were used for the evaluation of $(D_1^f)^2$ and $D_2^f$. Spin dilution was used for the $N_\tau=12$ ensembles and time-slice dilution for the $N_\tau=16$ ensembles. The diluted estimates entering $(D_1^f)^2$ were combined using unbiased estimators, while $D_2^f$ was evaluated without additional dilution.

For the physical light-quark mass, $m_l=m_s/27.4$, the analysis was extended to fourth order on the $N_\tau=12$ ensembles. In this case, 500 Gaussian random sources were used for $(D_1^f)^2$ together with time-slice dilution and unbiased estimators. All other traces required for fourth-order cumulants were also estimated with 500 Gaussian random sources, but without further dilution. To reduce the total number of inversions, we organized the trace measurements according to the tree-diagram prescription of Ref.~\cite{Gavai:2004sd}. {In the present calculation this prescription is used only to organize the required trace structures and reuse inversions efficiently. The traces themselves are evaluated with the MDWF Dirac operator, including the Pauli--Villars subtraction in Eq.~\eqref{eq:partitionfunc}; no staggered-specific determinant or operator identity is used.}

\subsection{Mass reweighting for $m_l=m_s/10$}

To correct for the mistuning of the line of constant physics in the $m_l=m_s/10$ ensembles, we apply configuration-by-configuration mass reweighting~\cite{Hasenfratz:2008fg,Liu:2012gm}.

\begin{figure}[!htbp]
    \includegraphics[width=0.36\linewidth]{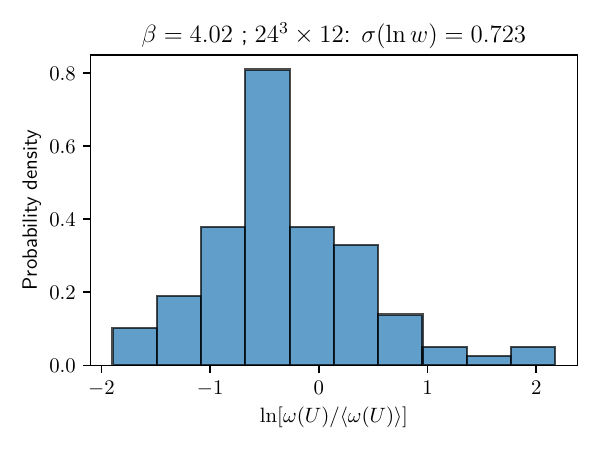}
    \includegraphics[width=0.36\linewidth]{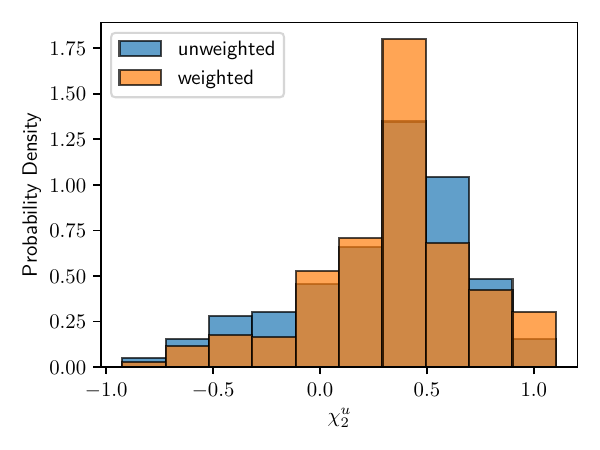}
    \includegraphics[width=0.36\linewidth]{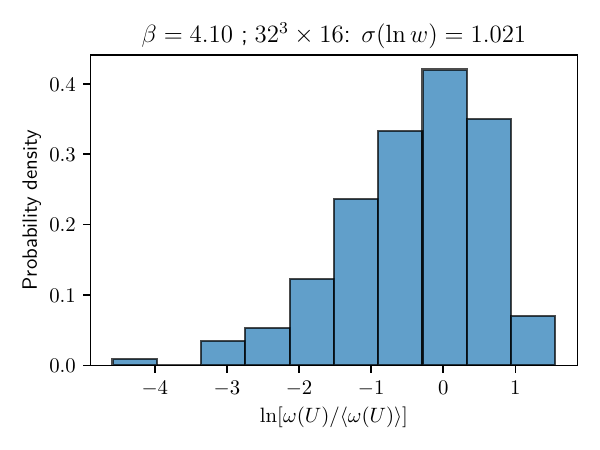}
    \includegraphics[width=0.36\linewidth]{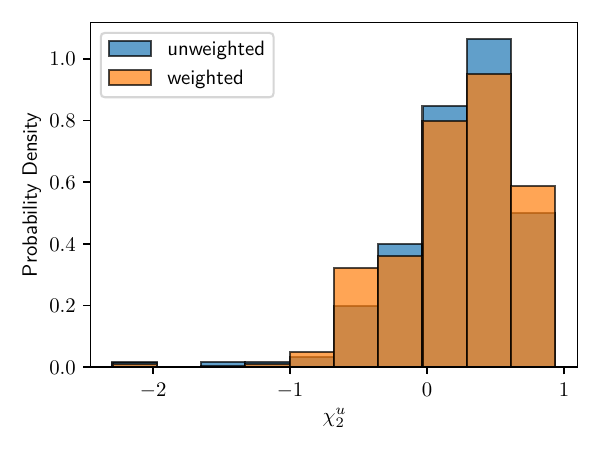}
    \caption{Distributions of the logarithm of the reweighting factor, $\ln(\omega)$, and of $\chi_2^u$ before and after reweighting for representative ensembles at $N_{\tau}=12$ (top) and $N_{\tau}=16$ (bottom).}
    \label{fig:reweighting-qns}
\end{figure}

For the $N_\tau=12$ ensembles, the largest mass shift occurs at $\beta=4.02$, where the input masses are changed from $(m_l^{\rm old},m_s^{\rm old})=(0.00051,0.05887)$ to $(m_l^{\rm new},m_s^{\rm new})=(0.001135,0.064565)$. The corresponding distributions of $\ln(\omega)$ and of $\chi_2^u$ before and after reweighting are shown in Fig.~\ref{fig:reweighting-qns}. For this largest reweighting step, we obtain $\sigma(\ln\omega)=0.73$, which corresponds to a typical weight range $\omega \in [e^{-0.73},e^{+0.73}] \approx [0.48,2.08]$ around the median scale.

As a complementary diagnostic, we estimate an effective sample-size fraction from the ratio of jackknife errors before and after reweighting,
\begin{equation}
\frac{\mathrm{ESS}_{\chi}}{N}
\equiv
\left(\frac{\sigma_{\rm before}^{\rm jack}}{\sigma_{\rm after}^{\rm jack}}\right)^2
=
\frac{1}{1.261^2}
\simeq 0.63.
\end{equation}
This indicates that about $63\%$ of the original ensemble is effectively retained for this observable after reweighting. The $\chi_2^u$ distribution is only mildly distorted, indicating that the reweighting remains under control within the present statistical precision.

For the $N_\tau=16$ ensembles, the largest mass shift occurs at $\beta=4.10$, where the input masses are changed from $(m_l^{\rm old},m_s^{\rm old})=(0.00484742,0.0484742)$ to $(m_l^{\rm new},m_s^{\rm new})=(0.003630,0.048205)$. In this case, we find $\sigma(\ln\omega)=1.021$, corresponding to $\omega \in [e^{-1.021},e^{+1.021}] \approx [0.36,2.78]$, and $\mathrm{ESS}_{\chi}/N \simeq 0.62$. As for the $N_\tau=12$ ensembles, the $\chi_2^u$ distribution changes only modestly after reweighting. Since this is the largest mass shift in the $N_\tau=16$ ensemble set, these diagnostics support the use of mass reweighting for the full $N_\tau=16$ dataset within current uncertainties.

\subsection{Results}

\begin{figure}[!htbp]
    \centering
    \includegraphics[width=0.30\textwidth]{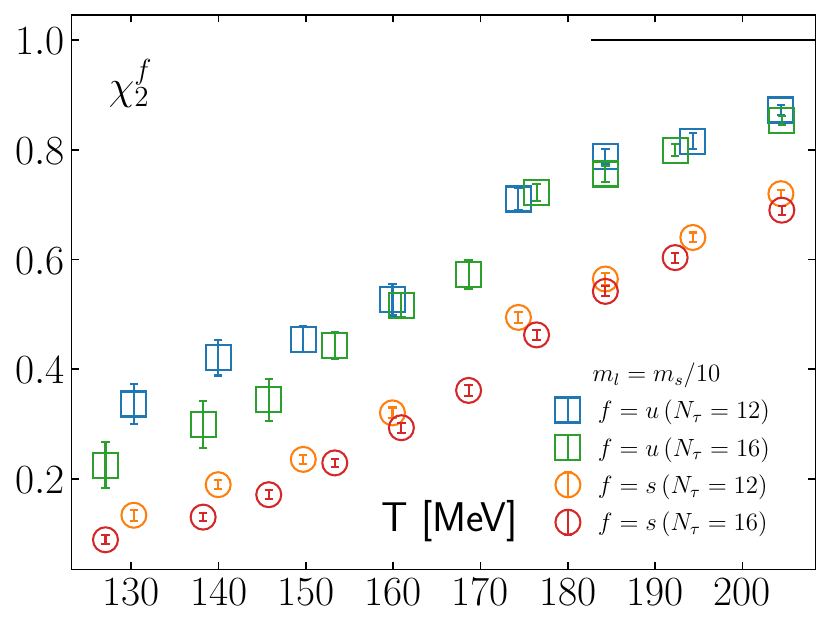}
    \hspace{1.5cm}
    \includegraphics[width=0.30\textwidth]{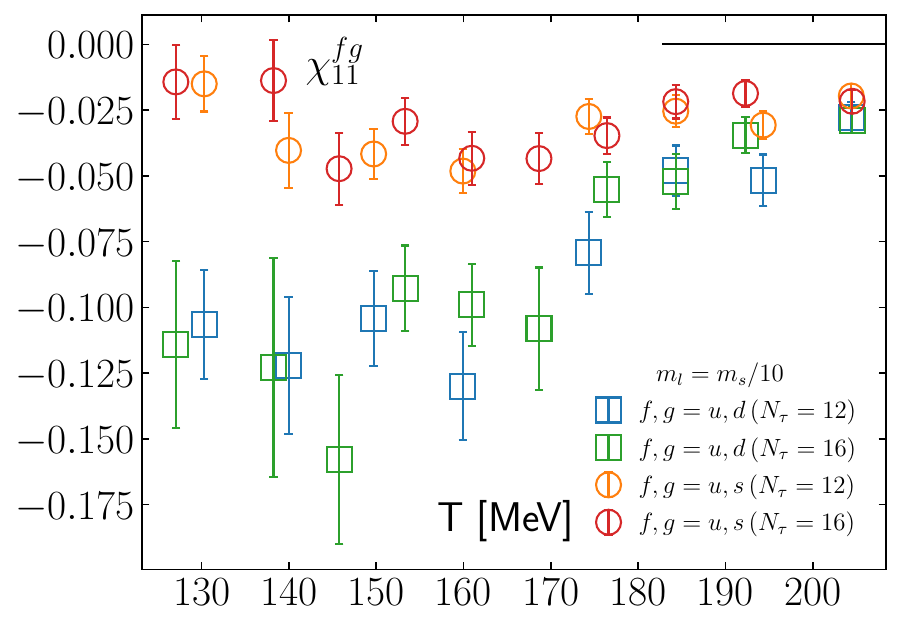}
    \includegraphics[width=0.30\textwidth]{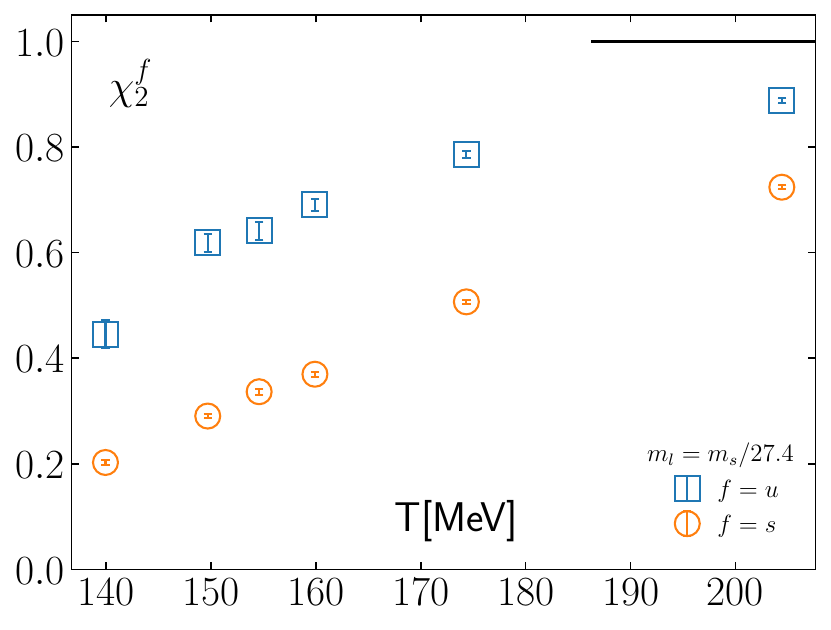}
    \hspace{1.5cm}
    \includegraphics[width=0.30\textwidth]{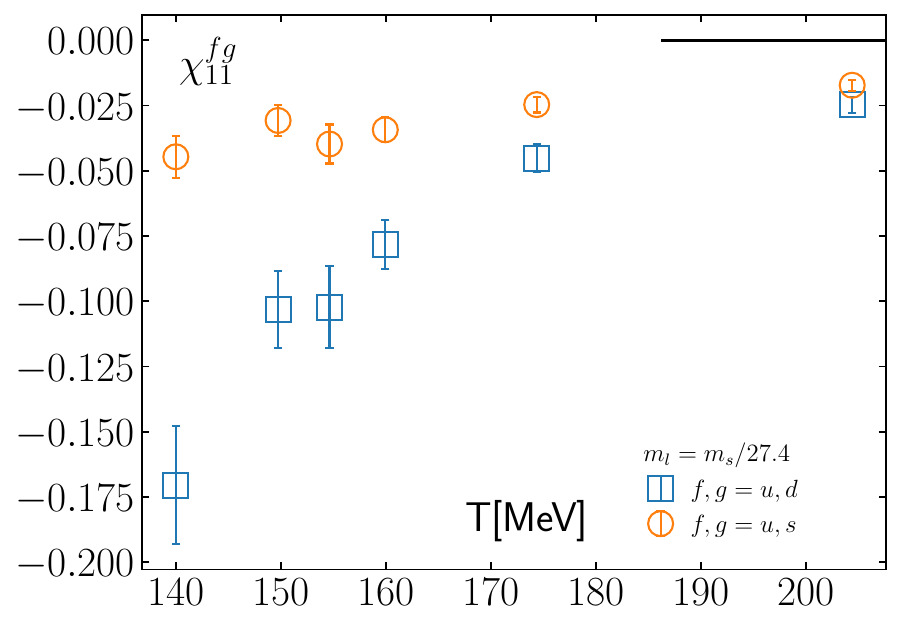}
    \caption{Diagonal quark number susceptibilities ($\chi_2^u,\chi_2^s$; left panels) and off-diagonal susceptibilities ($\chi_{11}^{ud},\chi_{11}^{us}$; right panels). The upper panels show results for the heavier light-quark mass, $m_l=m_s/10$, on lattices with $N_\tau=12$ and 16. The lower panels show results for the physical light-quark mass, $m_l=m_s/27.4$, on lattices with $N_\tau=12$. The black line denotes the continuum Stefan--Boltzmann limit for massless free fermions.}
    \label{fig:qns}
\end{figure}

Figure~\ref{fig:qns} shows the second-order quark number susceptibilities as functions of temperature for both light-quark masses. At low temperatures, all susceptibilities are small. They rise rapidly in the crossover region and gradually approach their Stefan--Boltzmann limits at higher temperatures. The diagonal susceptibilities are determined with significantly smaller errors than the off-diagonal ones, reflecting the larger disconnected contributions in $\chi_{11}^{ud}$ and $\chi_{11}^{us}$.

For $m_l=m_s/10$, the comparison between $N_\tau=12$ and 16 provides a first assessment of cutoff effects. Within the present uncertainties, the overall temperature dependence is similar on the two lattices, although some discretization effects remain visible in the crossover region. {The visible difference between the $N_\tau=12$ and 16 data near $T\simeq 140~\mathrm{MeV}$ motivated a blocking check. Appendix~\ref{app:resampling} compares the jackknife errors for the two relevant ensembles using block sizes of 100, 200, and 400 molecular-dynamics trajectories. The errors show no systematic growth with increasing block length, so the observed separation is not explained by an obvious underestimate of the statistical errors due to residual autocorrelations. We do not assign a continuum interpretation to this separation.} For $m_l=m_s/27.4$, the qualitative behavior is similar, but the transition region is shifted to lower temperature, consistent with the expected reduction of the pseudocritical temperature at smaller light-quark mass.

These results are qualitatively consistent with earlier studies using staggered fermions~\cite{Borsanyi:2011sw,Bazavov:2011nk}. 

\begin{figure}[!htbp]
\includegraphics[scale=0.30]{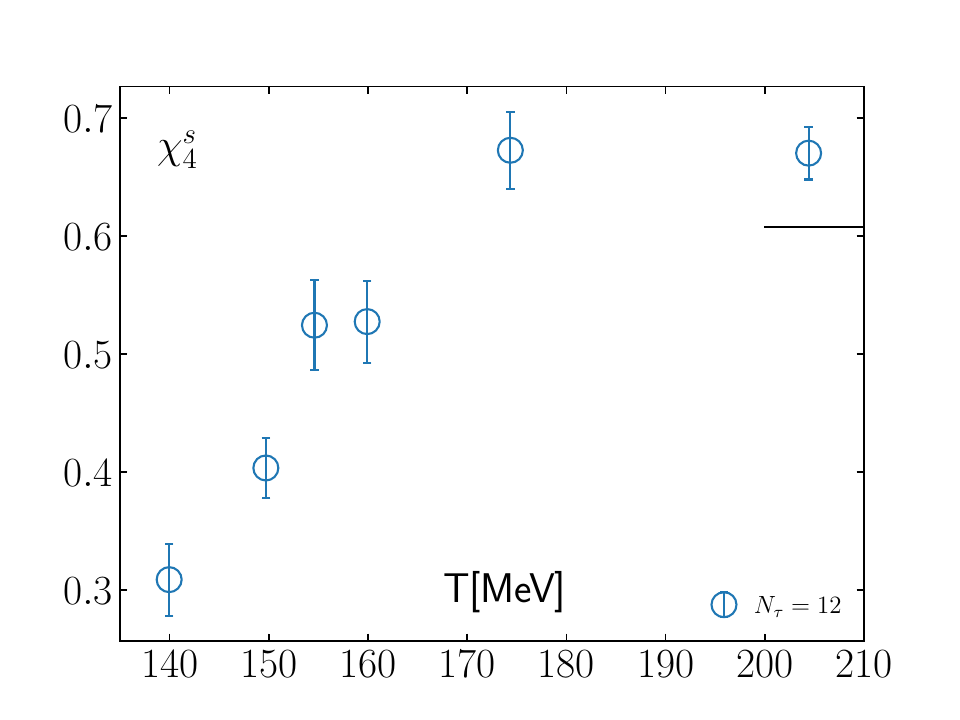}
\includegraphics[scale=0.30]{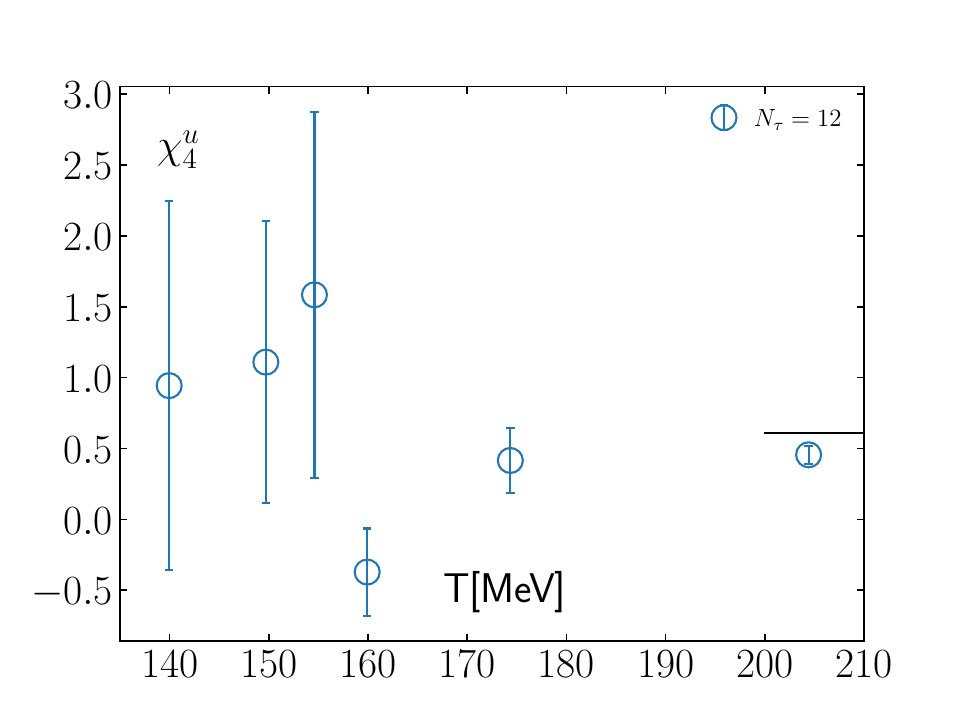}
\includegraphics[scale=0.30]{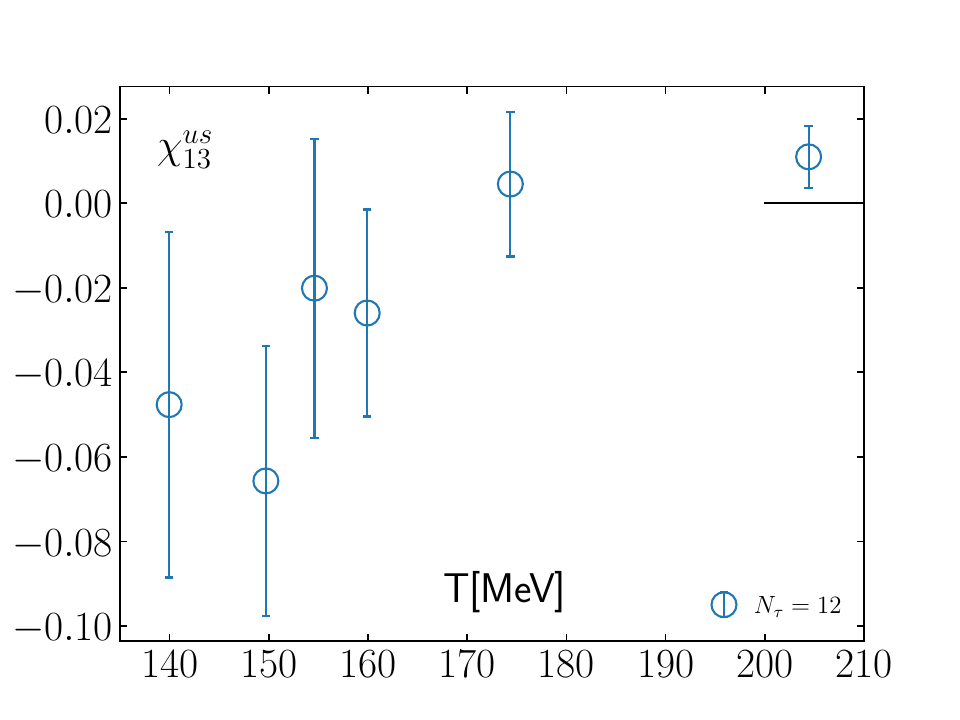}
\includegraphics[scale=0.30]{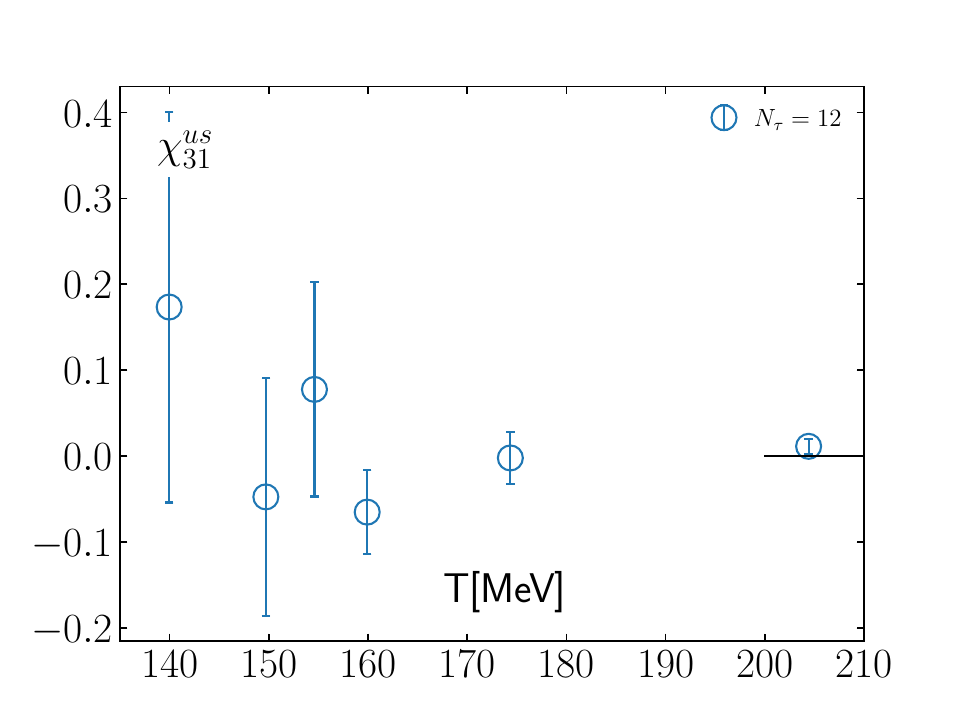}
\includegraphics[scale=0.30]{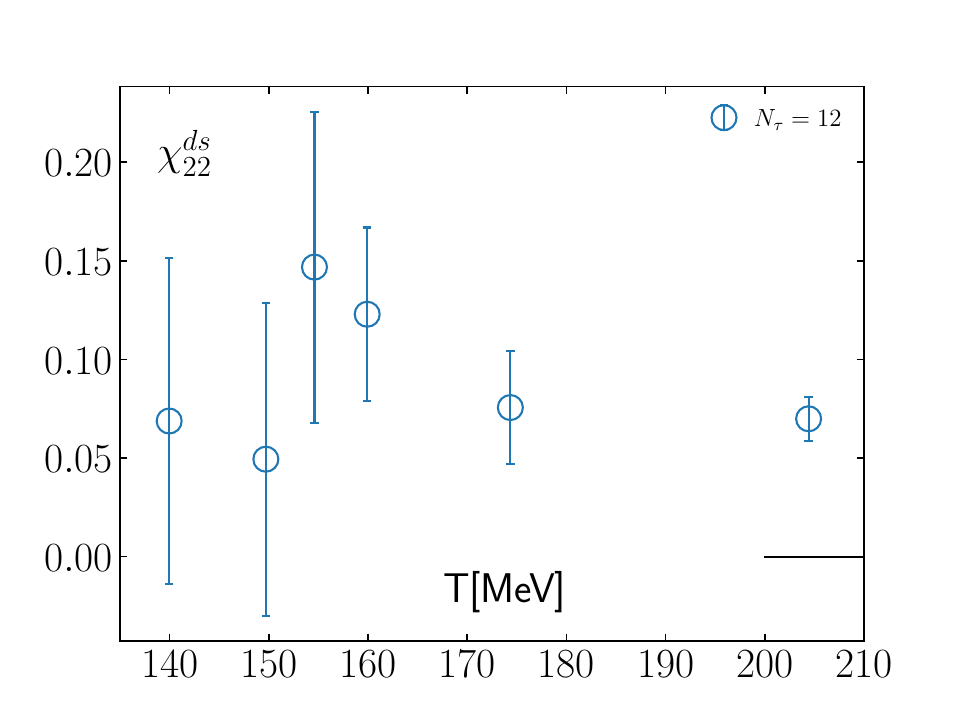}
\includegraphics[scale=0.30]{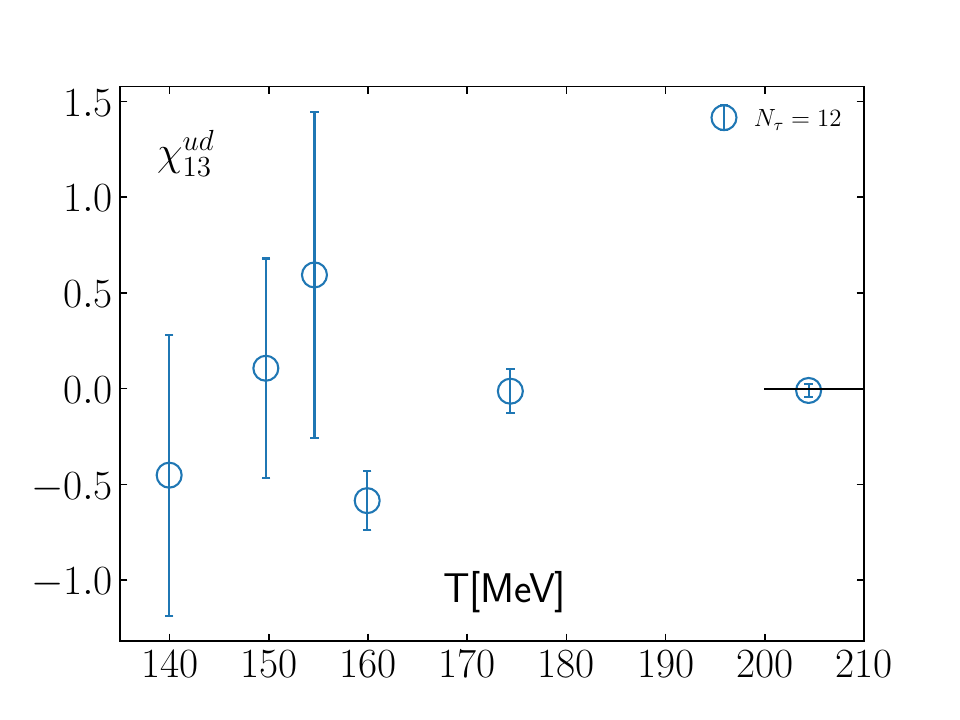}
\includegraphics[scale=0.30]{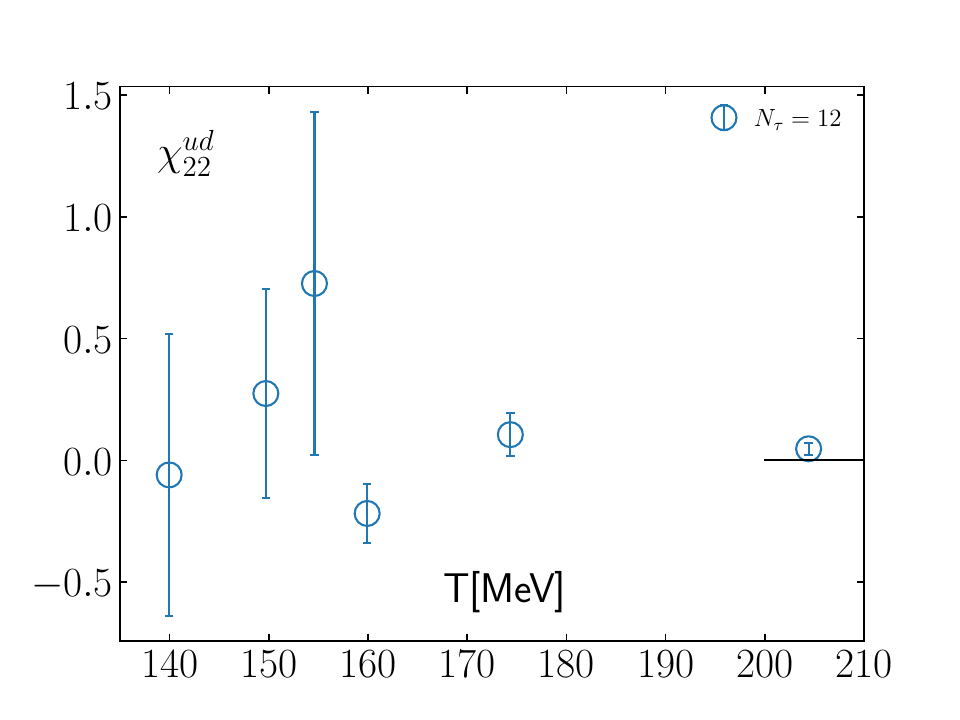}
\includegraphics[scale=0.30]{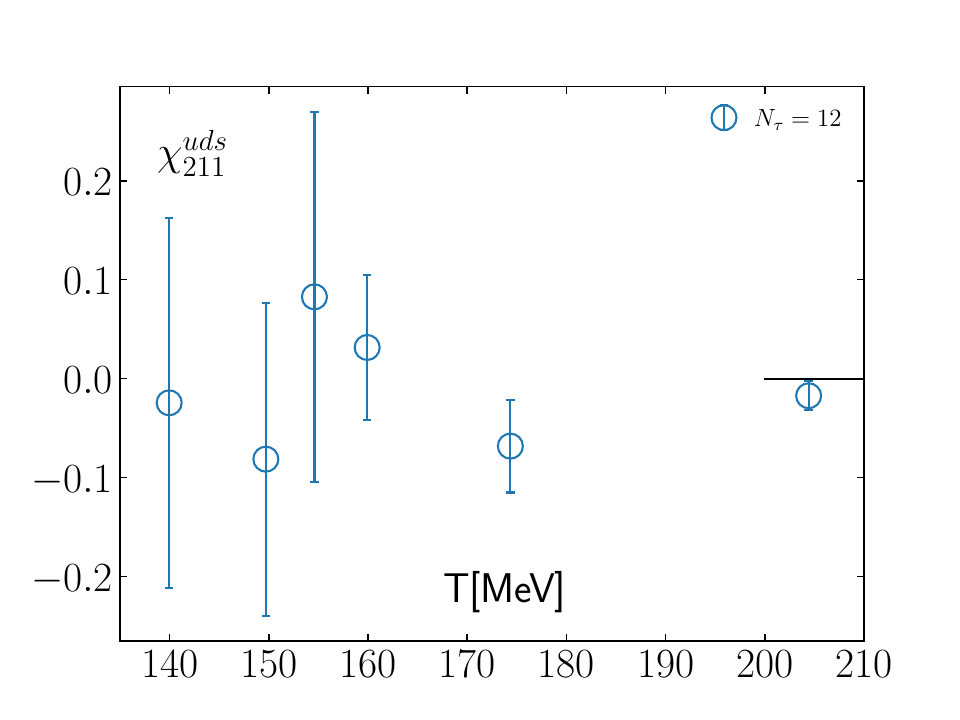}
\includegraphics[scale=0.30]{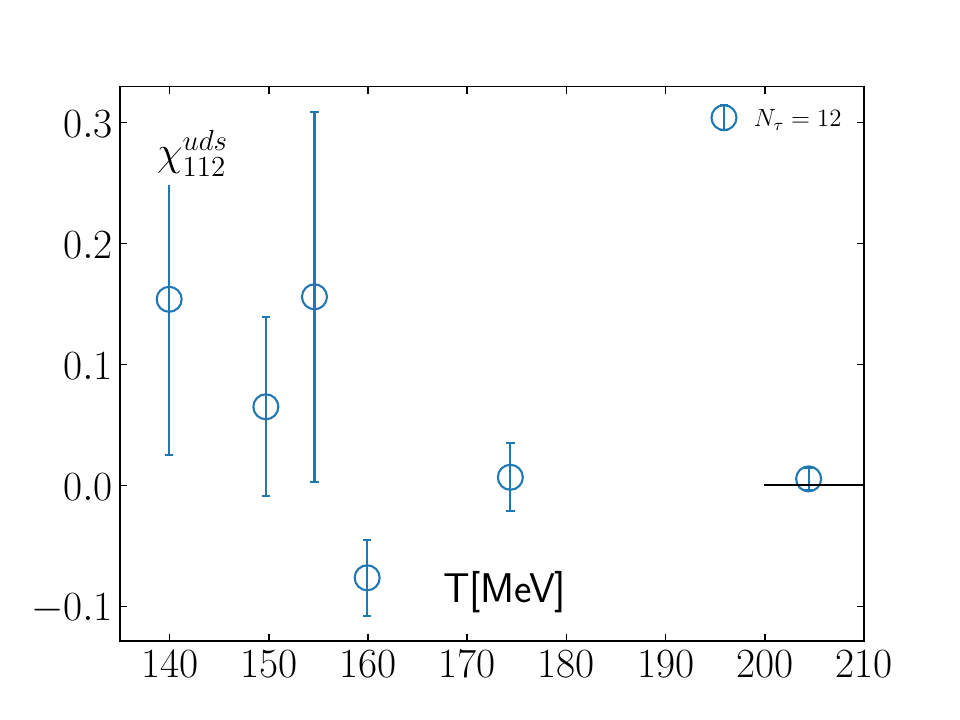}
\caption{Fourth-order quark number susceptibilities for $m_l=m_s/27.4$ on lattices with temporal extent $N_\tau=12$. The black line denotes the Stefan--Boltzmann limit of an ideal quark-gluon gas.}
\label{fig:fourth_order}
\end{figure}

Figure~\ref{fig:fourth_order} shows the fourth-order quark number susceptibilities for the physical light-quark mass. These observables enter the NNLO Taylor expansion of the pressure and are substantially more challenging numerically than the second-order susceptibilities. In particular, channels involving multiple derivatives with respect to the light-quark chemical potentials are affected by large disconnected contributions and correspondingly large statistical errors. Even with 500 Gaussian random sources and dilution, several of these observables remain only qualitatively resolved in the low-temperature region. {A representative blocking test for selected fourth-order quark-number susceptibilities is shown in Appendix~\ref{app:resampling}. The statistical errors remain stable within approximately 10--15\% over the block sizes considered, with no systematic increase with block length.} At higher temperatures, however, the statistical control improves and the data show the expected approach toward the free quark-gas limit.

In the next section, we use these quark number susceptibilities to construct fluctuations of conserved charges.
\section{Second-order conserved-charge fluctuations}
\label{sec:cnfresults}

We now turn to second-order fluctuations of the conserved charges baryon number ($B$), electric charge ($Q$), and strangeness ($S$) in $(2+1)$-flavor QCD. The analysis is performed for two light-quark masses, $m_l=m_s/10$ and $m_l=m_s/27.4$, the latter corresponding to the physical pion mass. Our main goal is to assess the light-quark-mass dependence of these observables and to compare the MDWF results with HRG expectations and with existing staggered-fermion calculations.

The conserved-charge chemical potentials are related to the quark chemical potentials through
\begin{equation}
\mu_u = \frac{1}{3}\mu_B + \frac{2}{3}\mu_Q,\qquad
\mu_d = \frac{1}{3}\mu_B - \frac{1}{3}\mu_Q,\qquad
\mu_s = \frac{1}{3}\mu_B - \frac{1}{3}\mu_Q - \mu_S.
\end{equation}
The pressure may therefore be expanded as
\begin{equation}
\frac{P}{T^4}
=
\sum_{i,j,k=0}^{\infty}
\frac{\chi_{ijk}^{BQS}}{i!\,j!\,k!}\,
\hat{\mu}_B^i \hat{\mu}_Q^j \hat{\mu}_S^k .
\label{eq:PdefinitionBQS}
\end{equation}
At second order, the conserved-charge susceptibilities are related to the quark-number susceptibilities by
\begin{align}
\chi_2^S &= \chi_2^s,
&
\chi_2^B &= \tfrac{1}{9}\bigl(2\chi_2^u + \chi_2^s + 2\chi_{11}^{ud} + 4\chi_{11}^{us}\bigr),
&
\chi_2^Q &= \tfrac{1}{9}\bigl(5\chi_2^u + \chi_2^s - 4\chi_{11}^{ud} - 2\chi_{11}^{us}\bigr),
\nonumber\\
\chi_{11}^{QS} &= \tfrac{1}{3}\bigl(\chi_2^s - \chi_{11}^{us}\bigr),
&
\chi_{11}^{BS} &= -\tfrac{1}{3}\bigl(\chi_2^s + 2\chi_{11}^{us}\bigr),
&
\chi_{11}^{BQ} &= \tfrac{1}{9}\bigl(\chi_2^u - \chi_2^s + \chi_{11}^{ud} - \chi_{11}^{us}\bigr).
\label{eq:second_order_cnf}
\end{align}
For degenerate light quarks, $m_u = m_d$, one has $\chi_2^u = \chi_2^d$ and $\chi_{11}^{us} = \chi_{11}^{ds}$, relations that have already been used above. The combinations appearing in Eq.~(\ref{eq:second_order_cnf}) also help explain the observed error pattern: in $\chi_2^Q$, disconnected contributions partially cancel, thereby reducing the statistical uncertainty, whereas in $\chi_2^B$ they add constructively, leading to a substantially noisier observable. Strange-sector observables are generally better controlled because of the heavier strange-quark mass.

For comparison, we also consider the noninteracting HRG model, in which the pressure is written as a sum over hadronic states,
\begin{eqnarray}
\frac{P}{T^4}
&=&
\sum\limits_{H \in \mathrm{baryons}}
\frac{g_H}{2\pi^2}
\left(\frac{m_H}{T}\right)^2
\sum\limits_{k=1}^{\infty}
\frac{(-1)^{k+1}}{k^2}
K_2\!\left(\frac{k m_H}{T}\right)
\exp\!\left[\frac{k\vec{C}_H \cdot \vec{\mu}}{T}\right]
\nonumber\\
&&+
\sum\limits_{H \in \mathrm{mesons}}
\frac{g_H}{2\pi^2}
\left(\frac{m_H}{T}\right)^2
\sum\limits_{k=1}^{\infty}
\frac{1}{k^2}
K_2\!\left(\frac{k m_H}{T}\right)
\exp\!\left[\frac{k\vec{C}_H \cdot \vec{\mu}}{T}\right],
\label{eq:ideal_gas_press}
\end{eqnarray}
where $g_H$, $m_H$, and $\vec{C}_H=(B,Q,S)$ denote the degeneracy, mass, and conserved-charge vector of hadron $H$, and $K_2$ is the modified Bessel function of the second kind. Generalized susceptibilities are obtained by differentiating Eq.~\eqref{eq:ideal_gas_press} with respect to the corresponding chemical potentials.

\subsection{Cutoff dependence at $m_l=m_s/10$}

\begin{figure*}[htbp]
  \centering
  \includegraphics[scale=0.34]{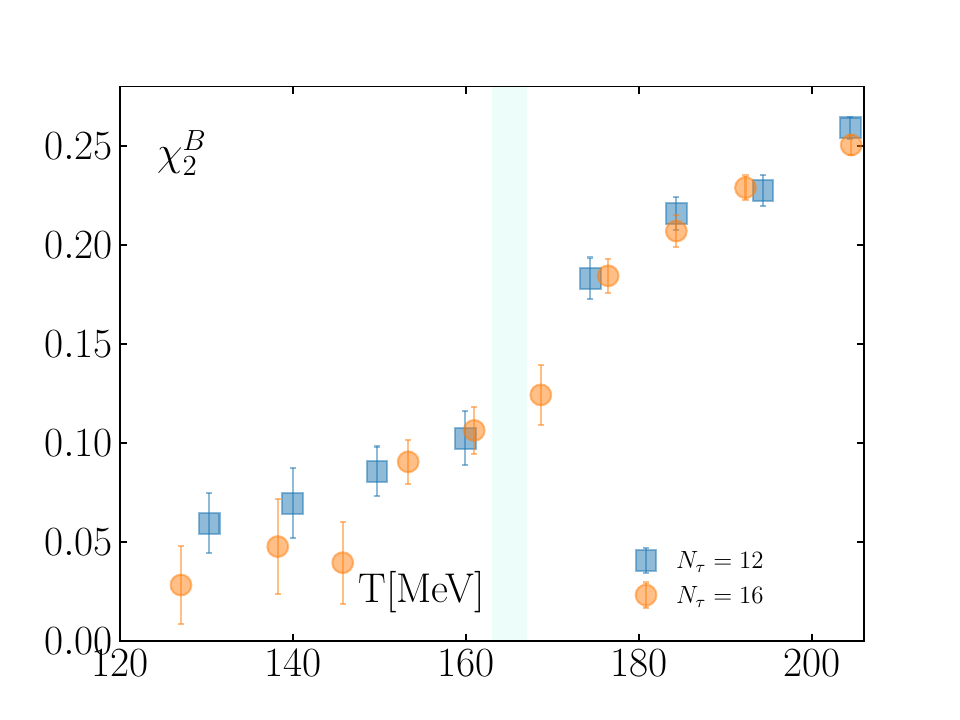}
  \includegraphics[scale=0.34]{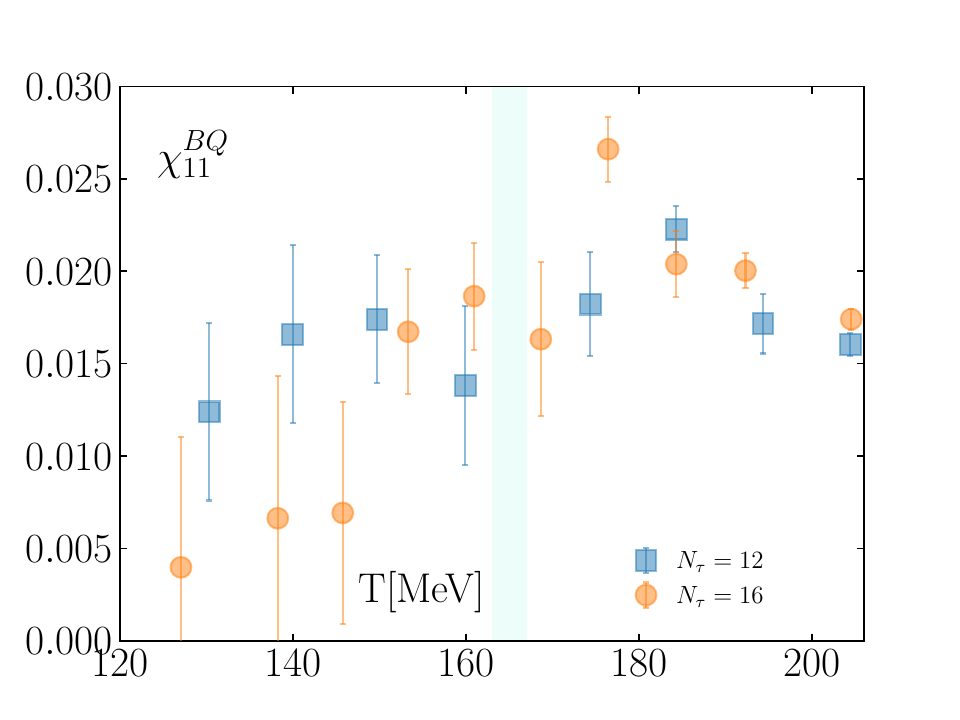}
  \includegraphics[scale=0.34]{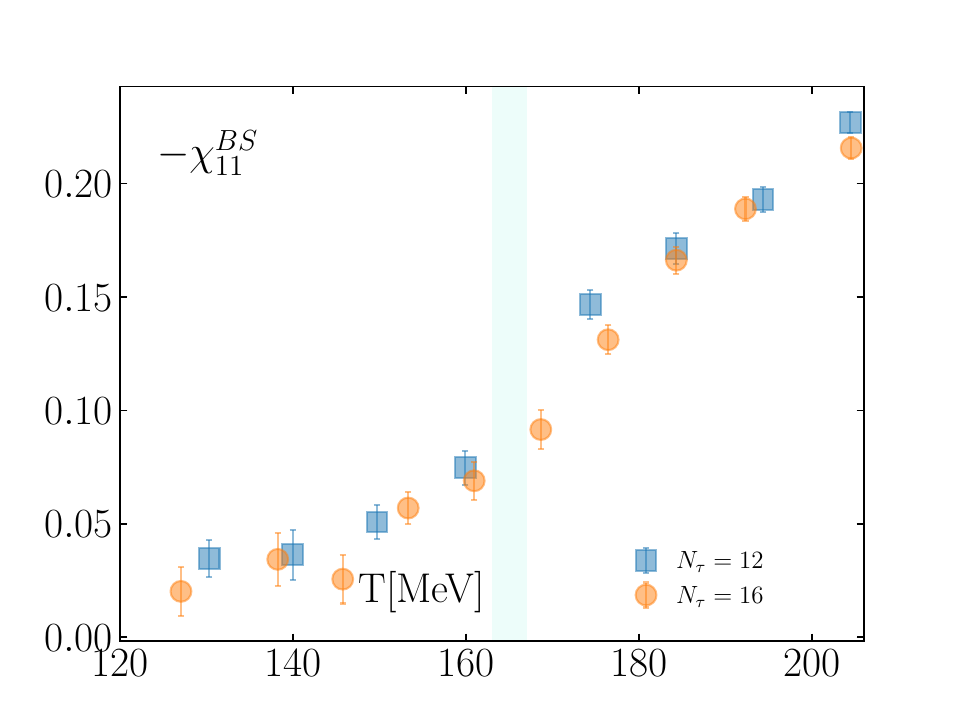}
  \includegraphics[scale=0.34]{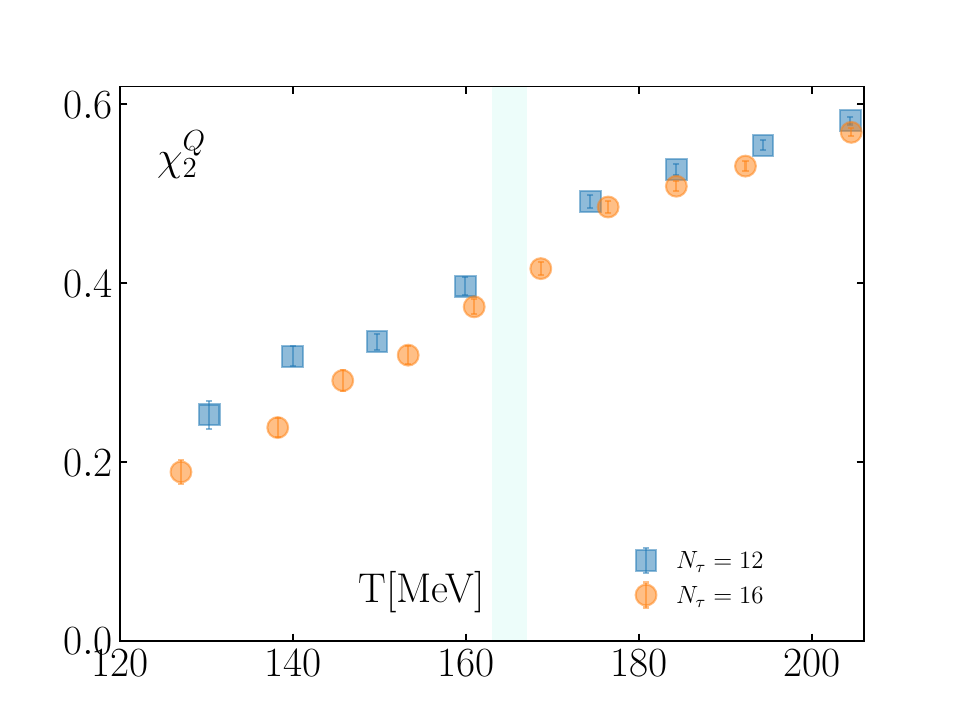}
  \includegraphics[scale=0.34]{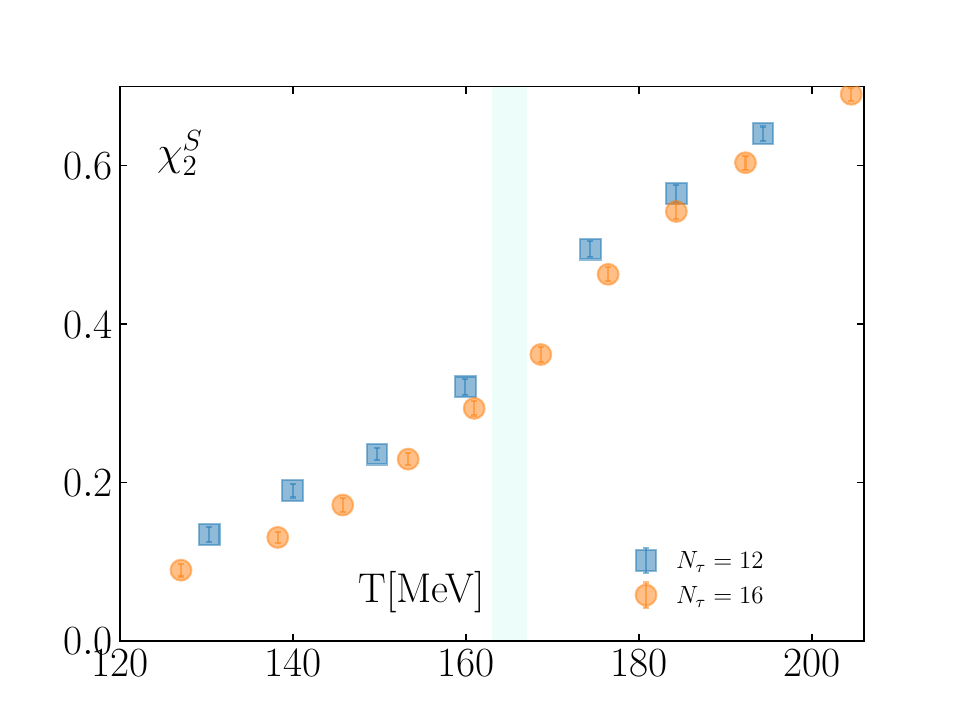}
  \includegraphics[scale=0.34]{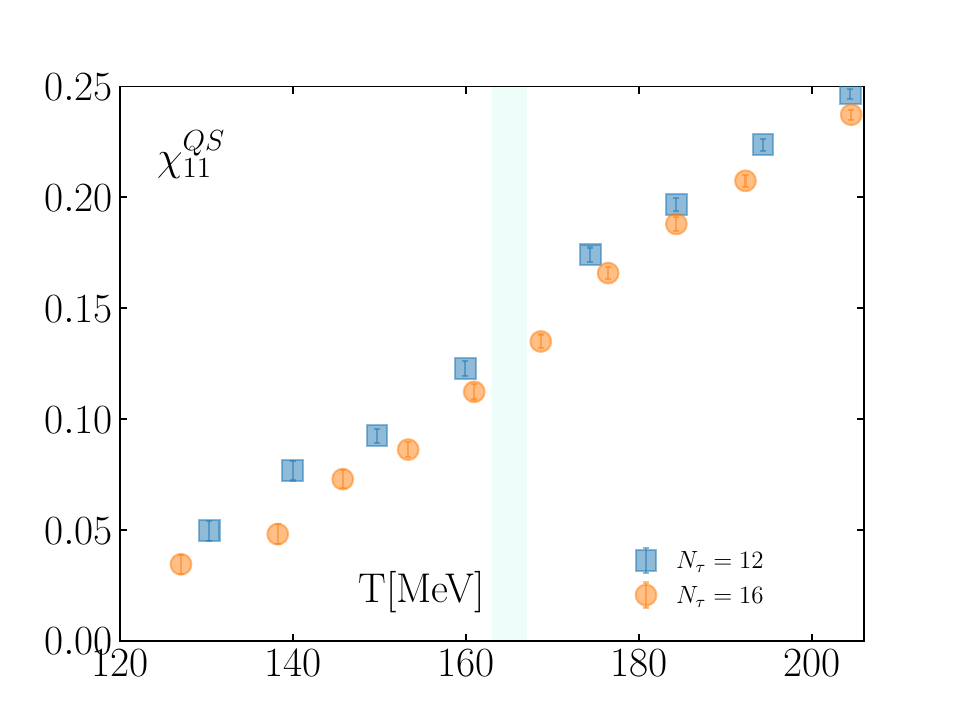}
  \caption{Second-order conserved-charge cumulants for $m_l=m_s/10$ as functions of temperature. The upper row shows observables with dominant baryonic contributions, while the lower row shows observables dominated by mesonic contributions. Results are shown for $N_\tau=12$ and 16 along the line of constant physics. {The shaded band marks the MDWF crossover range, $T_{\rm pc}=165(2)~\mathrm{MeV}$.}}
  \label{fig:01mssecondoffdiag}
\end{figure*}

We first examine the temperature dependence of second-order conserved-charge fluctuations for the heavier light-quark mass, $m_l=m_s/10$, using lattices with temporal extents $N_\tau=12$ and 16. Figure~\ref{fig:01mssecondoffdiag} shows baryon-dominated cumulants in the upper panels and meson-dominated cumulants in the lower panels. {In the following figures, shaded vertical bands mark the MDWF crossover ranges, $T_{\rm pc}=165(2)~\mathrm{MeV}$ for $m_l=m_s/10$ and $T_{\rm pc}=155(5)~\mathrm{MeV}$ for $m_l=m_s/27.4$, as applicable.}

Within the current statistical precision, the baryon-dominated cumulants $\chi_2^B$, $\chi_{11}^{BQ}$, and $\chi_{11}^{BS}$ show only small differences between $N_\tau=12$ and 16. The meson-dominated cumulants exhibit somewhat more visible cutoff dependence. In particular, the pion-sensitive electric-charge fluctuation $\chi_2^Q$ differs by less than about $4\%$ at the two lowest temperatures, whereas the kaon-dominated observables $\chi_2^S$ and $\chi_{11}^{QS}$ show effects at the level of about $1\%$. The mixed cumulant $\chi_{11}^{BQ}$ remains the noisiest quantity in this set. Above the crossover region, cutoff effects appear small within the present uncertainties.

\subsection{Comparison with staggered discretizations at the physical pion mass}

\begin{figure}[htbp]
  \centering
  \includegraphics[scale=0.34]{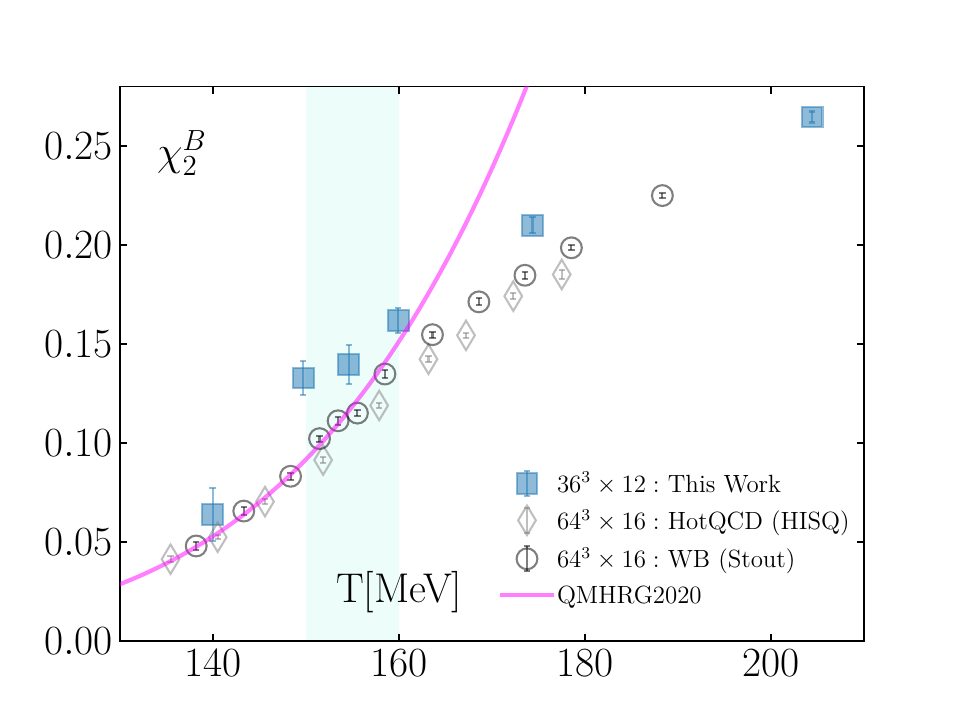}
  \includegraphics[scale=0.34]{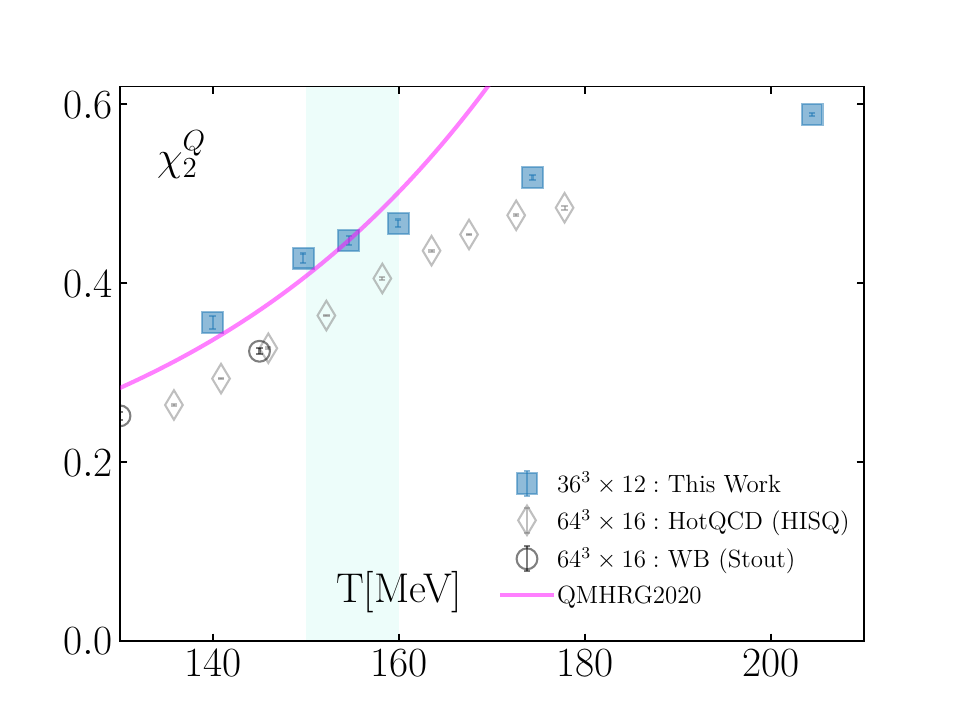}
  \includegraphics[scale=0.34]{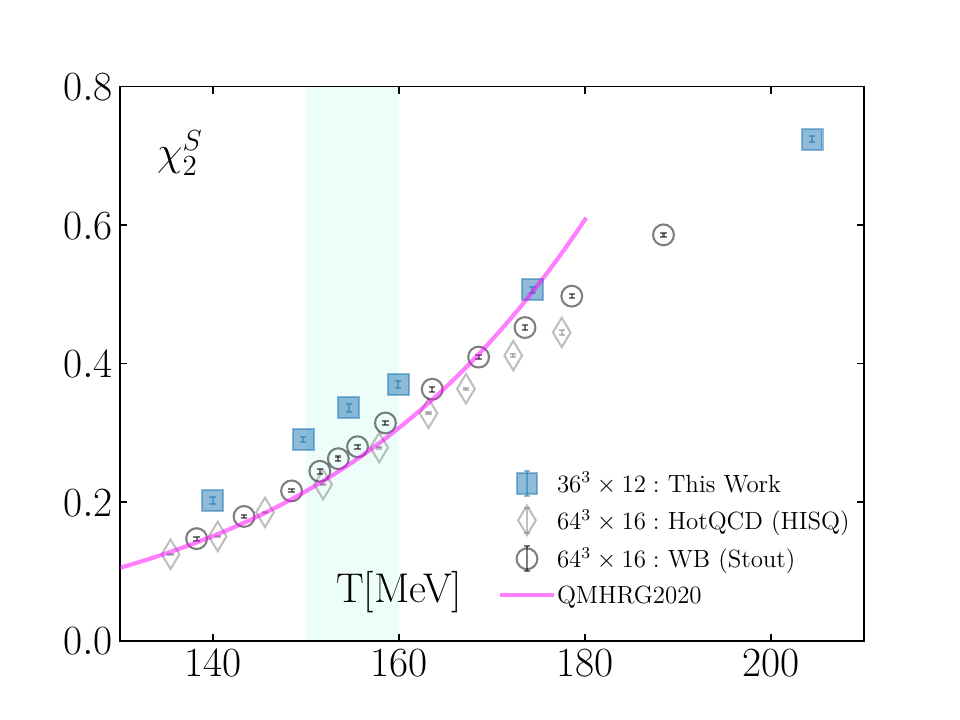}
  \includegraphics[scale=0.34]{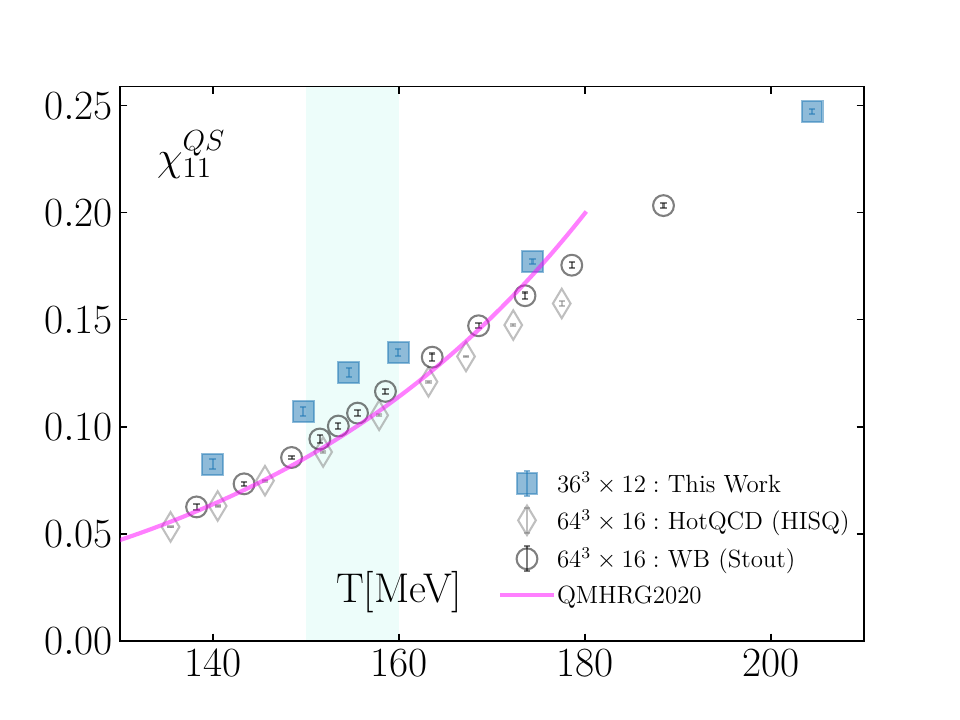}
  \includegraphics[scale=0.34]{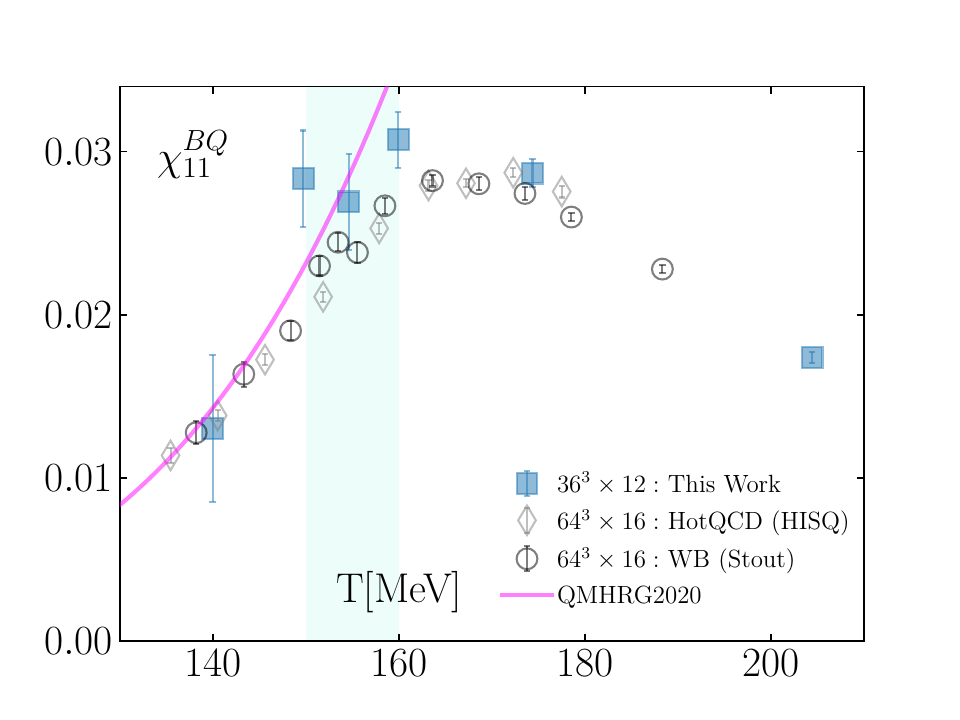}
  \includegraphics[scale=0.34]{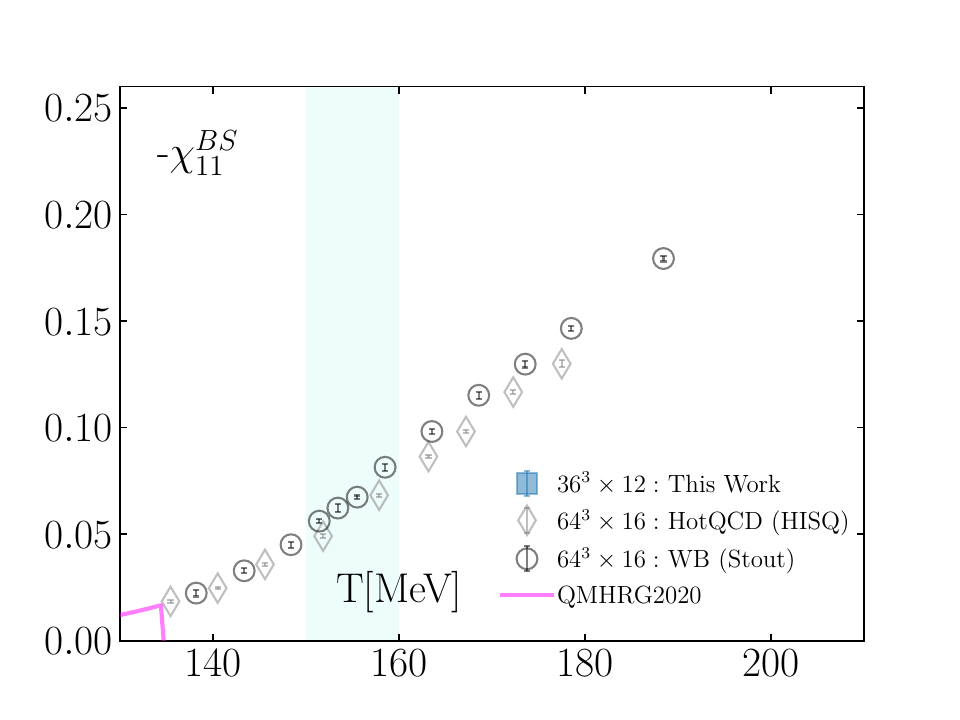}
  \caption{Second-order conserved-charge cumulants from MDWF at $m_l=m_s/27.4$, compared with HISQ results from HotQCD~\cite{Bollweg:2021vqf}, stout results from the Wuppertal--Budapest Collaboration~\cite{Bellwied:2015rza}, and QMHRG2020 calculations. {The shaded band marks the MDWF crossover range, $T_{\rm pc}=155(5)~\mathrm{MeV}$.}}
  \label{fig:mlphycssecondeleccharge}
\end{figure}

We next compare the MDWF results at the physical light-quark mass, $m_l=m_s/27.4$, with staggered-fermion calculations at finite lattice spacing. In Fig.~\ref{fig:mlphycssecondeleccharge}, we compare our data with results obtained using the HISQ action by the HotQCD Collaboration~\cite{Bollweg:2021vqf} and stout fermions by the Wuppertal--Budapest Collaboration~\cite{Bellwied:2015rza}. QMHRG2020 results are shown as a hadronic baseline.

At low temperatures, the largest visible difference between the MDWF and staggered results occurs in $\chi_2^Q$, which is particularly sensitive to the light pseudoscalar sector. Taste-symmetry breaking in staggered formulations increases the root-mean-square pion mass, $m_\pi^{\rm RMS}$~\cite{HotQCD:2012fhj}, and can therefore affect this channel more strongly. The ordering of the MDWF and staggered results is qualitatively consistent with this expectation, although the present data do not isolate taste-breaking effects quantitatively. For $\chi_2^S$, $\chi_{11}^{QS}$, and the baryon-related cumulants, the different calculations show broadly similar qualitative temperature dependence within present uncertainties. Since the physical-mass MDWF results are available at only one lattice spacing and one volume, and the calculations compared in Fig.~\ref{fig:mlphycssecondeleccharge} use different discretizations, lattice spacings, and volumes, this figure provides a finite-cutoff comparison between discretizations. Additional MDWF lattice spacings and volumes are required for quantitative conclusions.

\subsection{Sensitivity to the light-quark mass}

\begin{figure}[htbp]
\includegraphics[scale=0.34]{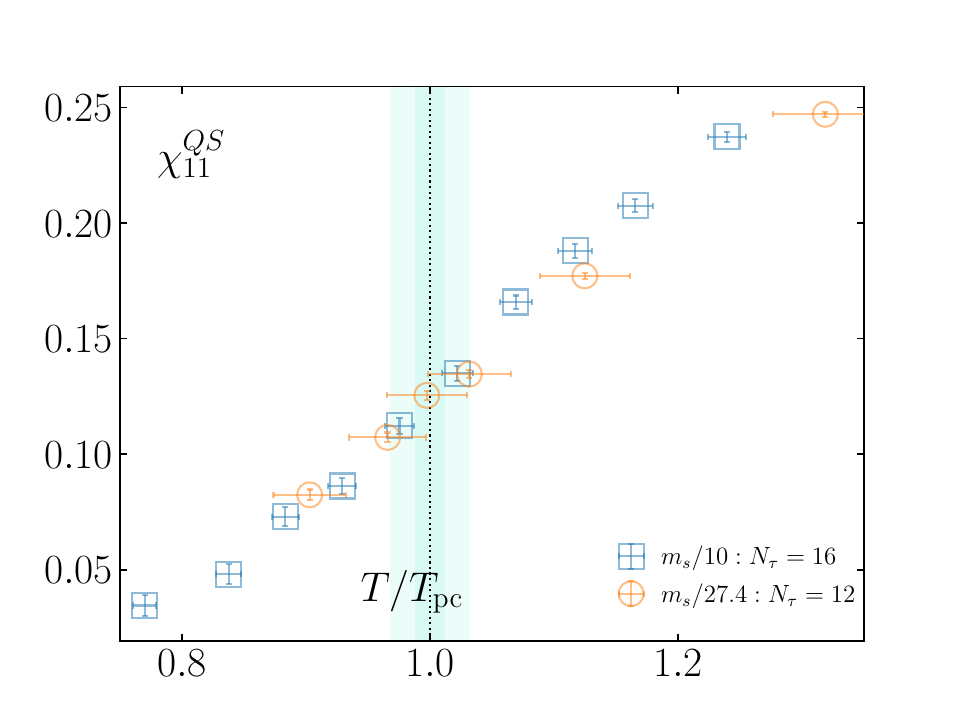}
\includegraphics[scale=0.34]{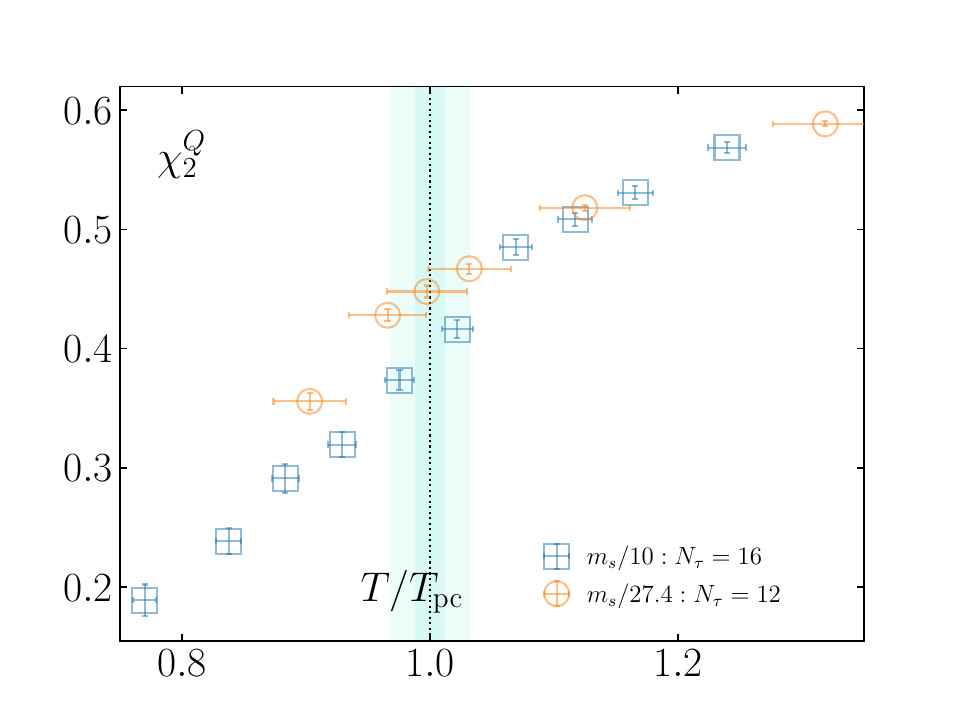}
\includegraphics[scale=0.34]{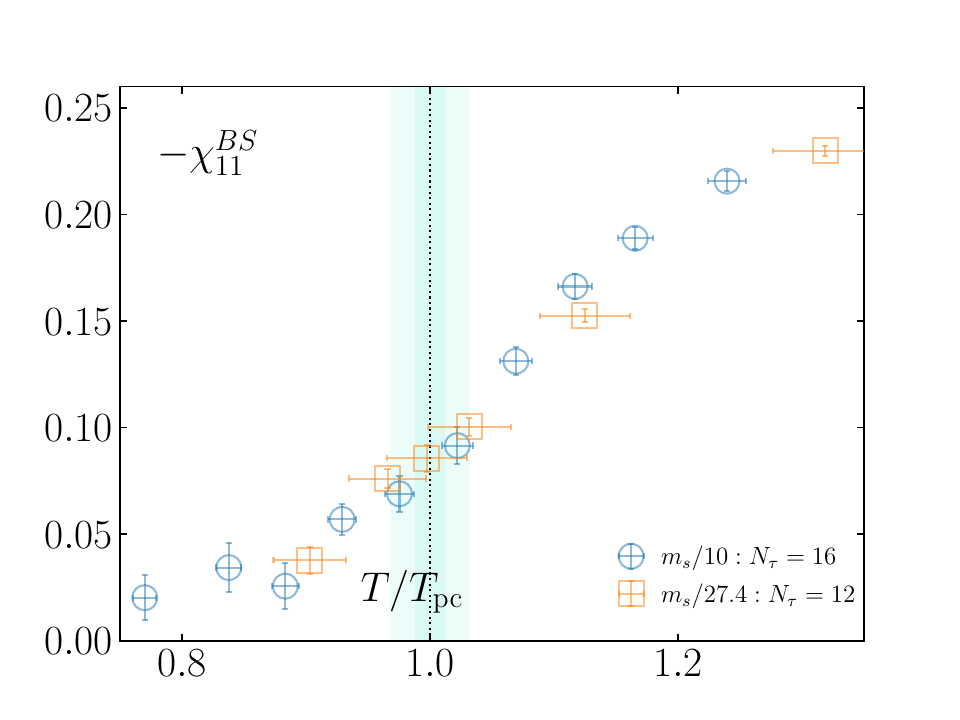}
\caption{Selected second-order conserved-charge cumulants, $\chi_{11}^{QS}$, $\chi_2^Q$, and $-\chi_{11}^{BS}$, for two light-quark masses, plotted as functions of $T/T_{\rm pc}$. We use $T_{\rm pc}=165(2)~\mathrm{MeV}$ for $m_l=m_s/10$ and $T_{\rm pc}=155(5)~\mathrm{MeV}$ for $m_l=m_s/27.4$. Horizontal error bars reflect the uncertainty in $T_{\rm pc}$, and the shaded bands indicate the corresponding crossover ranges around $T/T_{\rm pc}=1$.}
\label{fig:mlphycssecond}
\end{figure}

Figure~\ref{fig:mlphycssecond} compares $\chi_{11}^{QS}$, $\chi_2^Q$, and $-\chi_{11}^{BS}$ for $m_l=m_s/10$ and $m_l=m_s/27.4$ as functions of $T/T_{\rm pc}$. We use the corresponding MDWF pseudocritical temperatures, $T_{\rm pc}=165(2)~\mathrm{MeV}$ and $T_{\rm pc}=155(5)~\mathrm{MeV}$, respectively. Expressing the temperature in this form facilitates a comparison relative to the crossover scale but does not remove light-quark-mass effects.

	When expressed as functions of $T/T_{\rm pc}$, $\chi_{11}^{QS}$ and $-\chi_{11}^{BS}$ show little difference between the two masses within the present uncertainties. The main visible difference remains in the pion-dominated electric-charge fluctuation $\chi_2^Q$: the heavier-mass result lies below the physical-mass result below and around the crossover, while the two agree within uncertainties above the crossover. Much of the apparent light-quark-mass dependence in the strange-meson- and strange-baryon-dominated observables is therefore related to the shift in the pseudocritical temperature, while a clear residual dependence remains in the pion-sensitive electric-charge fluctuation.
The two datasets were obtained at different lattice spacings and volumes, so this comparison remains qualitative. A clean separation of light-quark-mass, cutoff, and finite-volume effects requires matched lattice spacings and additional volumes.
\section{Fourth-order conserved-charge fluctuations}
\label{sec:cnfresultsfourthorder}

Using Eq.~(\ref{eq:PdefinitionBQS}), the fourth-order conserved-charge cumulants can be expressed in terms of quark number susceptibilities as
\begin{align}
    \chi^{S}_{4} &= \chi^{s}_{4},
    \qquad
    \chi^{QS}_{13} = \frac{1}{3} (\chi^{s}_{4} -  \chi^{us}_{13}),
    \qquad
    \chi^{QS}_{22} = \frac{1}{9} (\chi^{s}_{4} - 2 \chi^{us}_{13} + 5 \chi^{us}_{22} - 4 \chi^{uds}_{112}), \\
    \chi^{QS}_{31} &= \frac{1}{27} \left(\chi^{s}_{4} - 7 \chi^{us}_{31} - 3 \chi^{us}_{13} + 15 \chi^{us}_{22} - 12 \chi^{uds}_{112} - 6 \chi^{uds}_{121} + 12 \chi^{uds}_{211}\right), \\
    \chi^{Q}_{4} &= \frac{1}{81} \Bigl(17 \chi^{d}_{4} + \chi^{s}_{4} - 32 \chi^{ud}_{31} - 28 \chi^{us}_{31} - 8\chi^{ud}_{13} - 4 \chi^{us}_{13} + 24 \chi^{ud}_{22} + 30 \chi^{us}_{22}
    \nonumber\\
    &\hspace{1.5cm}
    - 24 \chi^{uds}_{112} - 24 \chi^{uds}_{121} + 48 \chi^{uds}_{211} \Bigr), \\
    \chi^{BS}_{13} &= - \frac{1}{3} \left(\chi^{s}_{4} + 2 \chi^{ds}_{13} \right),
    \qquad
    \chi^{BQS}_{112} = \frac{1}{9} \left ( -\chi^{s}_{4} -  \chi^{us}_{13} + \chi^{us}_{22} + \chi^{uds}_{112}\right), \\
    \chi^{BQS}_{121} &= \frac{1}{27} \left(-\chi^{s}_{4} - 5\chi^{us}_{31} - 3 \chi^{us}_{22} + 3 \chi^{uds}_{121} + 6 \chi^{uds}_{112} \right), \\
    \chi^{BQ}_{13} &= \frac{1}{81} \Bigl(7\chi^{u}_{4} - \chi^{s}_{4}  + \chi^{us}_{13} - 4 \chi^{ud}_{31} - 8 \chi^{us}_{31}  - 12 \chi^{us}_{22} - 6 \chi^{ud}_{22} + 15 \chi^{ud}_{13}
    \nonumber\\
    &\hspace{1.5cm}
    - 12 \chi^{uds}_{211} + 15 \chi^{uds}_{112} + 15 \chi^{uds}_{121}\Bigr), \\
    \chi^{BS}_{22} &= \frac{1}{9} (\chi^{s}_{4} + 2 \chi^{us}_{22} + 4 \chi^{us}_{13} + 2 \chi^{uds}_{112}),
    \qquad
    \chi^{BQS}_{211} = \frac{1}{27} (\chi^{s}_{4} -  \chi^{us}_{31}  + 3 \chi^{ds}_{13} - 3 \chi^{uds}_{211}), \\
    \chi^{BQ}_{22} &= \frac{1}{81} \Bigl(5\chi^{u}_{4} + \chi^{s}_{4} - 2 \chi^{ud}_{13} + 2 \chi^{us}_{13} - 3 \chi^{ud}_{22} + 3\chi^{us}_{22} + 4 \chi^{ud}_{31} + 8 \chi^{us}_{31}
    \nonumber\\
    &\hspace{1.5cm}
    - 6 \chi^{uds}_{211} - 6 \chi^{uds}_{112} - 6 \chi^{uds}_{121} \Bigr), \\
    \chi^{BS}_{31} &= \frac{1}{27} (-\chi^{s}_{4} - 6 \chi^{us}_{13} - 6 \chi^{us}_{22} - 2 \chi^{us}_{31} - 6 \chi^{uds}_{112} - 3 \chi^{uds}_{211} - 3 \chi^{uds}_{121}), \\
    \chi^{BQ}_{31} &=  \frac{1}{81} \Bigl( \chi^{u}_{4} - \chi^{s}_{4} - 5 \chi^{us}_{13} + \chi^{us}_{31} - \chi^{ud}_{13} + 3 \chi^{ud}_{22} - 3 \chi^{us}_{22} + 5 \chi^{ud}_{31}
    \nonumber\\
    &\hspace{1.5cm}
    - 3 \chi^{uds}_{112} - 3 \chi^{uds}_{121} + 6 \chi^{uds}_{211} \Bigr), \\
    \chi^{B}_{4} &= \frac{1}{81} \Bigl(2\chi^{u}_{4} + \chi^{s}_{4} + 12 \chi^{us}_{22} + 6 \chi^{ud}_{22} + 8 \chi^{us}_{13} + 4 \chi^{ud}_{31} + 8 \chi^{us}_{31} + 4 \chi^{ud}_{13}
    \nonumber\\
    &\hspace{1.5cm}
    + 12 \chi^{uds}_{211} + 12 \chi^{uds}_{112} + 12 \chi^{uds}_{121} \Bigr).
\end{align}
In the following, we focus on selected fourth-order conserved-charge cumulants for which statistically meaningful trends can be identified. As discussed above, observables involving $\chi_4^u$ and/or several off-diagonal light-quark susceptibilities are numerically the most demanding in the hadronic regime, whereas strange-sector quantities are generally better resolved. In the following sections, we present fourth-order conserved-charge fluctuations calculated on the $m_l = m_s/27.4$, $N_\tau = 12$ lattice.

\subsection{Strangeness fluctuations and electric-charge--strangeness correlations}

\begin{figure}[htbp]
\includegraphics[scale=0.36]{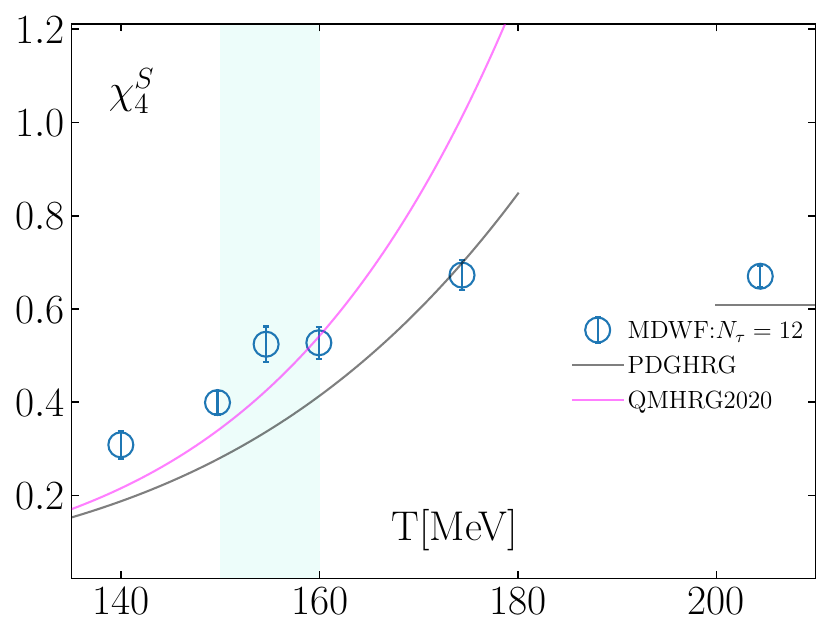}
\includegraphics[scale=0.36]{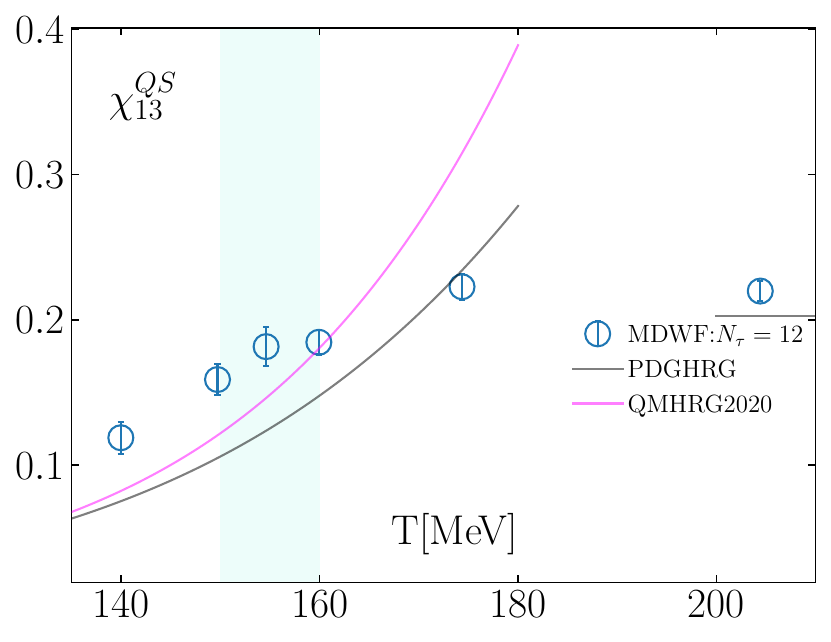}
\includegraphics[scale=0.36]{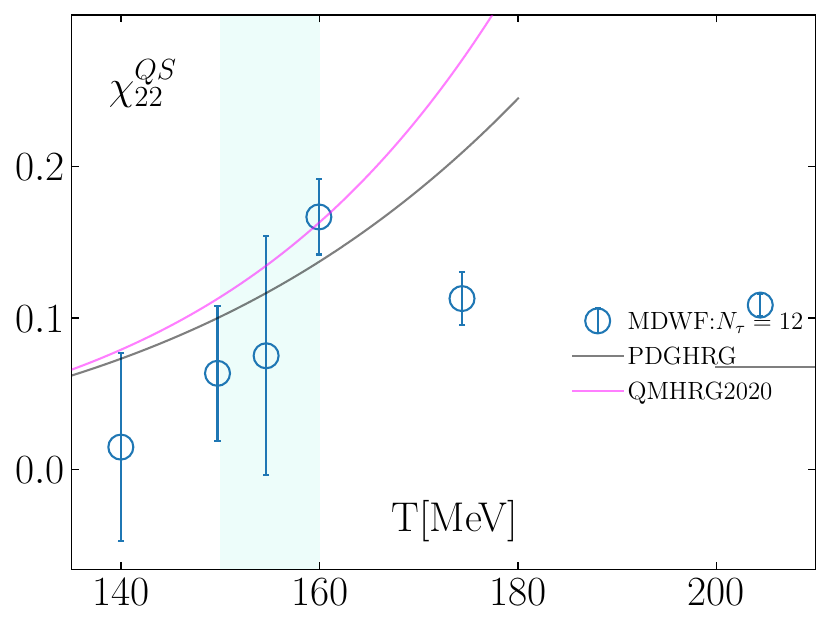}
\caption{Fourth-order strangeness fluctuations and electric-charge--strangeness correlations as functions of temperature. Results are compared with PDG-HRG and QMHRG2020 calculations. In the hadronic regime, these observables are primarily sensitive to the strange hadron spectrum, with charged kaons giving important contributions to the electric-charge--strangeness correlations. The black line denotes the Stefan--Boltzmann limit of an ideal quark-gluon gas. {The shaded band marks the MDWF crossover range, $T_{\rm pc}=155(5)~\mathrm{MeV}$.}}
\label{fig:fourth_order_kaonstrangeness}
\end{figure}

Figure~\ref{fig:fourth_order_kaonstrangeness} shows $\chi_4^S$, $\chi_{13}^{QS}$, and $\chi_{22}^{QS}$. Among these observables, $\chi_{22}^{QS}$ carries the largest statistical uncertainty, consistent with its stronger sensitivity to light-quark derivatives and the associated stochastic noise. {This also explains why the statistical error is not identical for $\chi_{13}^{QS}$ and $\chi_{22}^{QS}$. From the conserved-charge combinations written above, $\chi_{13}^{QS}$ is dominated by $\chi_4^s$ with one light--strange correction, whereas $\chi_{22}^{QS}$ contains several additional off-diagonal and mixed light--strange trace combinations. These subleading terms are numerically noisy and can enter through cancellations, so they affect the uncertainty of $\chi_{22}^{QS}$ more visibly.}

In the hadronic regime, these cumulants receive contributions from strange hadrons. The low-temperature behavior of the electric-charge--strangeness correlations is strongly influenced by charged strange mesons, in particular kaons, while strange baryons become increasingly relevant as the temperature approaches the crossover region. For $\chi_4^S$ and $\chi_{13}^{QS}$, the PDG-HRG baseline tends to lie below the lattice data at low temperature, whereas the QMHRG2020 calculation is generally closer to the observed trend. This pattern is consistent with sensitivity to additional strange hadronic states included in the QMHRG2020 spectrum.

Within a noninteracting HRG picture, $\chi_4^S$ receives contributions from both charged and neutral strange hadrons, while $\chi_{13}^{QS}$ and $\chi_{22}^{QS}$ are more strongly weighted toward electrically charged strange states. At low temperature this makes charged kaons particularly important. The difference between $\chi_{13}^{QS}$ and $\chi_{22}^{QS}$ below and around $T_{\rm pc}$ can also receive contributions from strange baryons, including multiply strange states such as the $\Xi$ and $\Omega$, although a quantitative decomposition would require higher statistics.

Above the crossover region, all three observables deviate from HRG calculations and show the expected trend toward quark-gluon degrees of freedom.

\subsection{Baryon-strangeness and mixed baryon-charge-strangeness correlations}

\begin{figure}[htbp]
\includegraphics[scale=0.34]{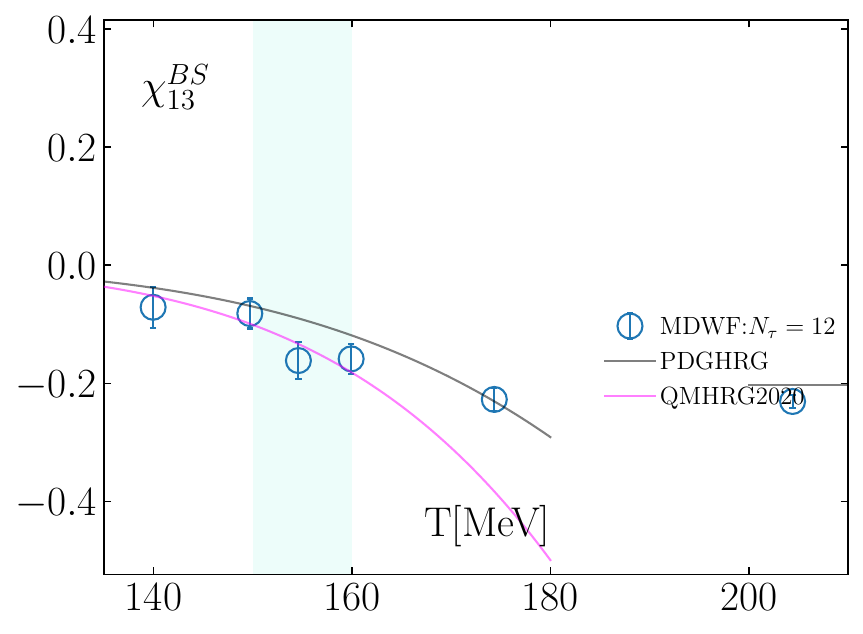}
\includegraphics[scale=0.34]{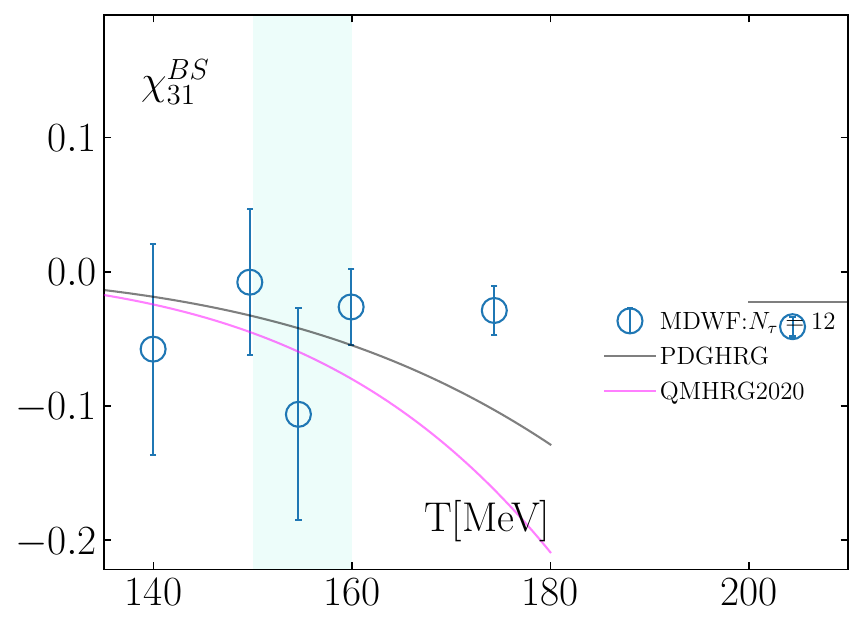}
\includegraphics[scale=0.34]{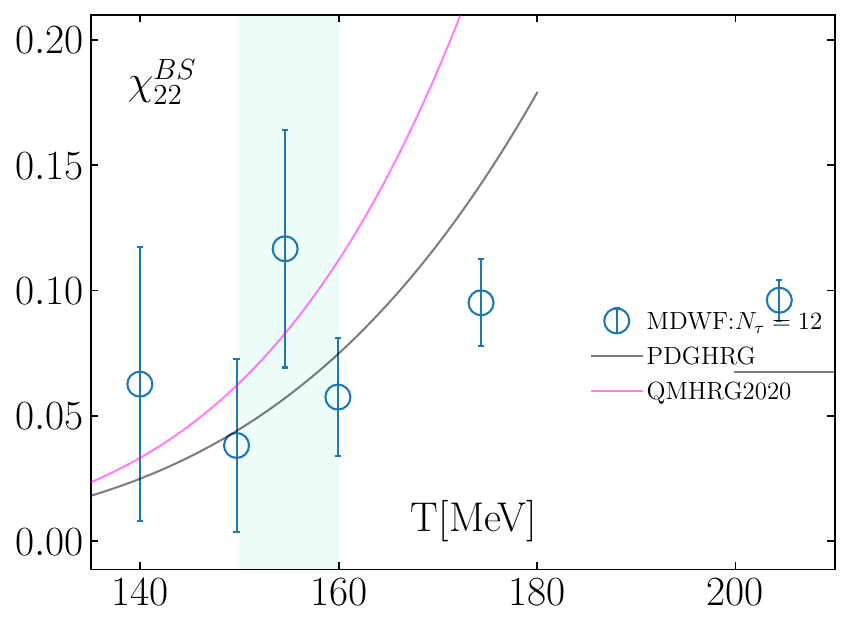}
\includegraphics[scale=0.34]{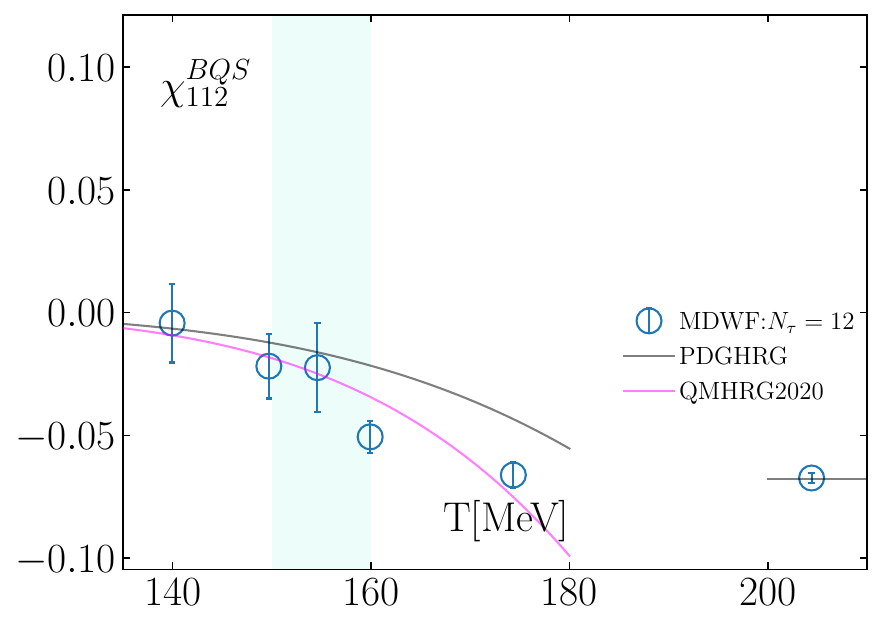}
\includegraphics[scale=0.34]{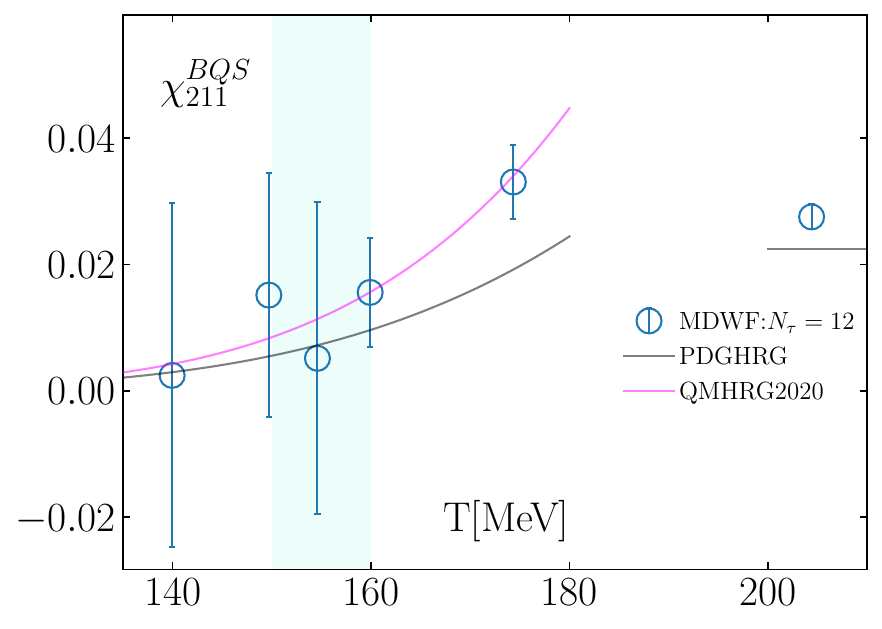}
\includegraphics[scale=0.34]{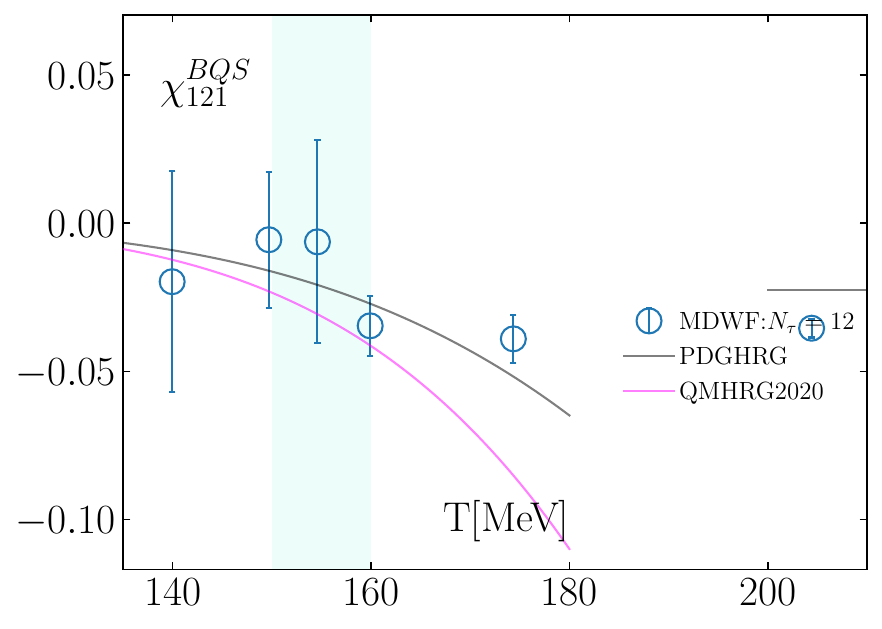}
\caption{Fourth-order baryon--strangeness and mixed baryon--charge--strangeness correlations as functions of temperature. Results are compared with PDG-HRG and QMHRG2020 calculations. In an HRG description, these observables receive contributions only from baryonic states; because they also contain strangeness derivatives, they are primarily sensitive to strange baryons. The black line denotes the Stefan--Boltzmann limit of an ideal quark-gluon gas. {The shaded band marks the MDWF crossover range, $T_{\rm pc}=155(5)~\mathrm{MeV}$.}}
\label{fig:fourth_order_baryon_strangeness}
\end{figure}

Figure~\ref{fig:fourth_order_baryon_strangeness} shows fourth-order baryon--strangeness and mixed baryon--charge--strangeness correlations. Since each observable contains at least one derivative with respect to $\mu_B$, mesons do not contribute in a noninteracting HRG description. These cumulants therefore isolate the baryonic sector; the additional strangeness derivatives make them particularly sensitive to strange baryons such as the $\Lambda$, $\Sigma$, $\Xi$, and $\Omega$.

For $\chi_{13}^{BS}$, the PDG-HRG result tends to lie below the lattice data near the pseudocritical temperature, while the QMHRG2020 calculation lies closer to the lattice trend. This behavior is consistent with sensitivity to additional strange baryonic states beyond the PDG-HRG baseline. As the temperature approaches $T_{\rm pc}$, however, both HRG descriptions begin to deviate from the lattice data. For the remaining observables shown in Fig.~\ref{fig:fourth_order_baryon_strangeness}, the lattice results are broadly consistent with the HRG calculations within present statistical uncertainties.

As expected, observables involving higher-order off-diagonal light- and strange-quark trace structures carry larger statistical errors. At higher temperatures, the data tend toward their Stefan--Boltzmann limits.

\subsection{Electric-charge and baryon--electric-charge fluctuations}

\begin{figure}[!htbp]
\includegraphics[scale=0.44]{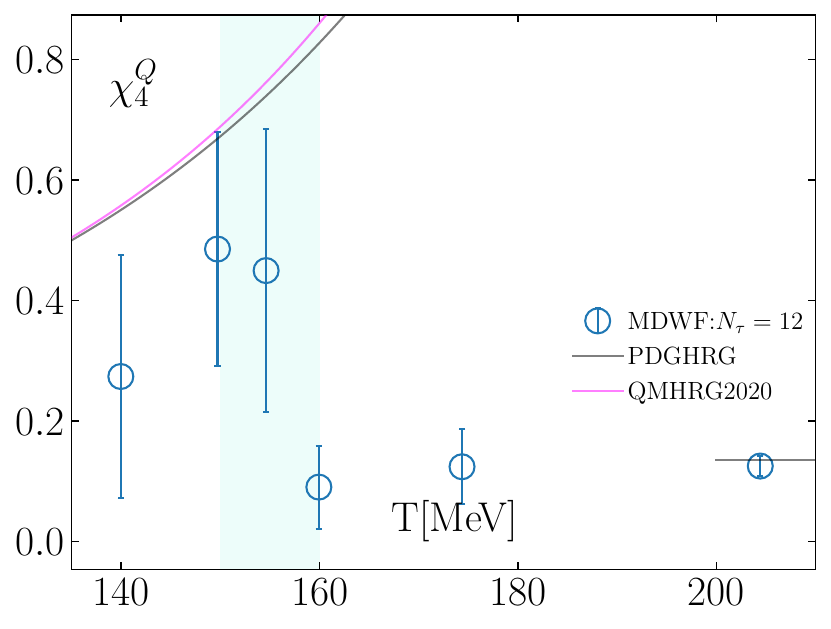}
\includegraphics[scale=0.44]{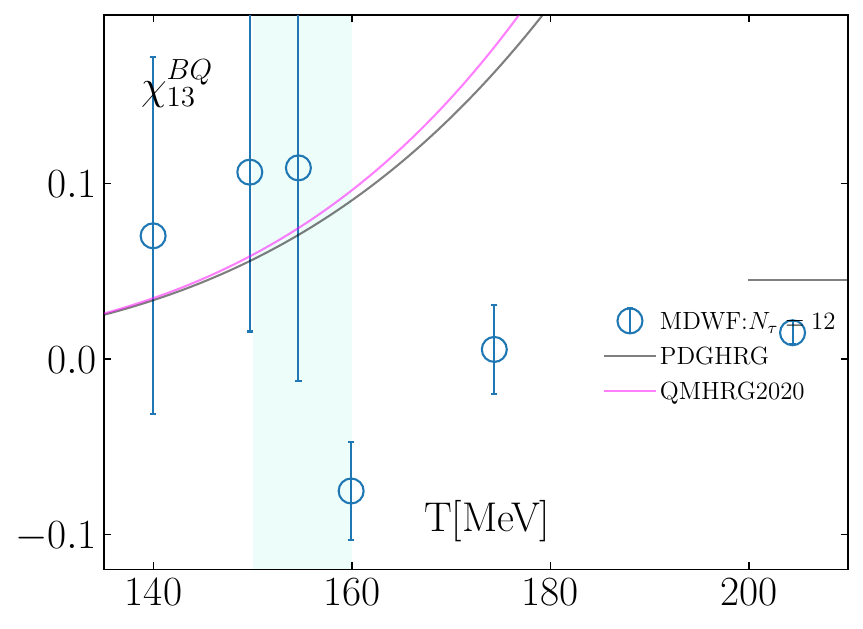}
\caption{Fourth-order electric-charge fluctuations and baryon--electric-charge correlations as functions of temperature. Results are compared with PDG-HRG and QMHRG2020 calculations. {The shaded band marks the MDWF crossover range, $T_{\rm pc}=155(5)~\mathrm{MeV}$.}}
\label{fig:fourth_order_elcharge}
\end{figure}

Figure~\ref{fig:fourth_order_elcharge} shows $\chi_4^Q$ and $\chi_{13}^{BQ}$. These observables involve third- and fourth-order derivatives with respect to the light-quark chemical potentials and are therefore among the noisiest quantities in our analysis. Accordingly, the discussion in this channel remains qualitative.

Within the present uncertainties, $\chi_4^Q$ appears to deviate from both PDG-HRG and QMHRG2020 in the hadronic regime. Since this observable is strongly influenced by the light-meson sector, the discrepancy may reflect interaction effects, limitations of the hadron-spectrum input, or both. We also observe that $\chi_{13}^{BQ}$ changes sign above the pseudocritical region. However, additional lattice spacings, larger volumes, and improved statistics will be required before this behavior can be interpreted more quantitatively. By contrast, $\chi_4^B$ is currently too noisy in the low-temperature regime to allow a useful quantitative discussion.

\subsection{Ratios of fourth- to second-order electric-charge and strangeness fluctuations}
\label{sec:cnfratiocomp}

The ratio of fourth- to second-order conserved-charge cumulants,
\[
R_{42}^X(\hat{\mu}_B,T) \equiv \frac{\chi_4^X}{\chi_2^X} + \mathcal{O}(\hat{\mu}_B^2),
\qquad X=Q,S,
\]
is often discussed as a useful diagnostic of the underlying degrees of freedom and may provide phenomenological input for freeze-out studies. In this work, we consider the leading-order ratios for electric charge and strangeness at vanishing baryon chemical potential.

\begin{figure}[htbp]
    \centering
    \includegraphics[scale=0.44]{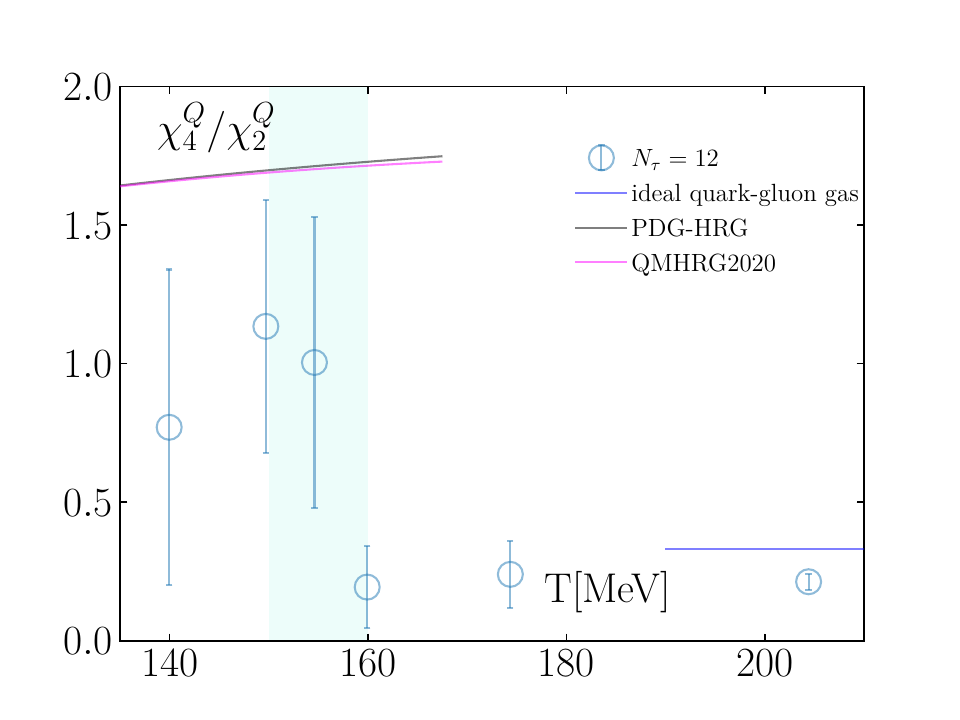}
    \includegraphics[scale=0.44]{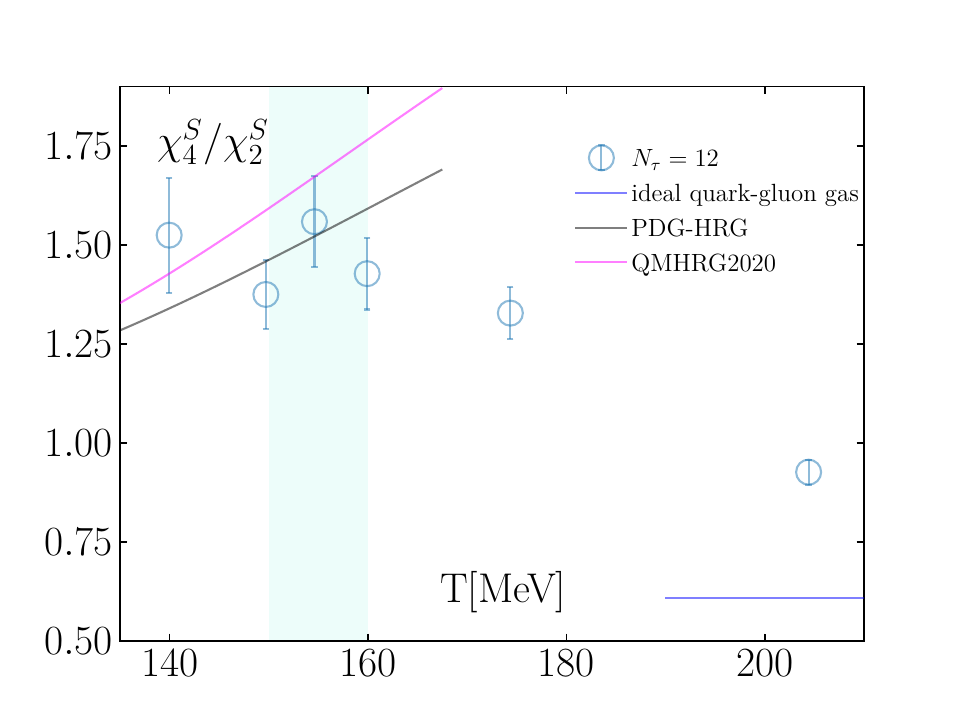}
    \caption{Leading-order kurtosis ratios for electric charge (left) and strangeness (right) as functions of temperature. Results are compared with PDG-HRG and QMHRG2020 calculations. The blue line denotes the Stefan--Boltzmann limit of an ideal quark-gluon gas. {The shaded band marks the MDWF crossover range, $T_{\rm pc}=155(5)~\mathrm{MeV}$.}}
    \label{fig:fourth_second_order_ratio}
\end{figure}

The temperature dependence of
\[
R_{42}^Q=\frac{\chi_4^Q}{\chi_2^Q},
\qquad
R_{42}^S=\frac{\chi_4^S}{\chi_2^S},
\]
obtained using MDWF, is shown in Fig.~\ref{fig:fourth_second_order_ratio}. At low temperatures, $R_{42}^Q$ lies below both HRG baselines within the present uncertainties, while $R_{42}^S$ is consistent with PDG-HRG and QMHRG2020. At higher temperatures, both ratios show the expected trend toward their Stefan--Boltzmann limits.

Near the pseudocritical region, the electric-charge ratio remains only weakly constrained,
\[
\frac{\chi_4^Q}{\chi_2^Q} =
\begin{cases}
1.05 \pm 0.45 & T = 149.7~\mathrm{MeV}, \\
1.00 \pm 0.52 & T = 154.6~\mathrm{MeV},
\end{cases}
\]
so that the present discussion of $R_{42}^Q$ is necessarily qualitative. By contrast, the strangeness ratio is better resolved:
\[
\frac{\chi_4^S}{\chi_2^S} =
\begin{cases}
1.38 \pm 0.10 & T = 149.7~\mathrm{MeV}, \\
1.55 \pm 0.12 & T = 154.6~\mathrm{MeV}.
\end{cases}
\]
This pattern is consistent with the generally better statistical control of strange-sector observables already seen in the individual cumulants.

\section{Summary and Outlook}

We have computed conserved-charge fluctuations in $(2+1)$-flavor QCD using MDWF along a line of constant physics. Two sets of ensembles were analyzed: $m_l/m_s=1/10$ at $N_\tau=12$ and 16, which allow an assessment of cutoff effects at heavier pion mass, and $m_l/m_s=1/27.4$ at $N_\tau=12$, which gives access to the light-quark-mass dependence down to the physical pion mass.

The second-order conserved-charge fluctuations show the expected rapid variation through the crossover. Below $T_{\rm pc}$, several channels are qualitatively compatible with HRG expectations within present uncertainties. When the two light-quark masses are compared as functions of $T/T_{\rm pc}$, the strange-meson- and strange-baryon-dominated observables largely follow a common trend, while $\chi_2^Q$ retains the clearest residual difference below and around the crossover. This behavior is qualitatively consistent with the stronger sensitivity of $\chi_2^Q$ to the light pseudoscalar sector. Since the physical-mass data are presently available at only one lattice spacing and one volume, no quantitative conclusion about discretization effects or agreement with HRG can be drawn.

We also computed selected fourth-order conserved-charge cumulants at the physical pion mass. These observables are numerically much more demanding, especially in channels involving several derivatives with respect to the light-quark chemical potentials and multiple disconnected contributions. Nevertheless, statistically meaningful trends can be identified in several strange-sector and mixed conserved-charge cumulants. In these channels, the QMHRG2020 baseline is often closer to the lattice data than the PDG-HRG baseline, while deviations from both HRG models become more visible across the crossover region. For electric-charge-related fourth-order observables, the current discussion remains largely qualitative because of the larger statistical uncertainties.

Overall, our results show that MDWF can be used effectively for fluctuation studies at finite temperature. Future work should focus on improving the statistical precision, extending the calculations to additional lattice spacings and larger volumes, and performing controlled continuum extrapolations, especially at the physical pion mass. Such extensions are needed to establish robust lattice-QCD baselines for conserved-charge fluctuations relevant to heavy-ion phenomenology.

\section*{Acknowledgments}
The project is supported by the MEXT as ``Program for Promoting Researches on the Supercomputer Fugaku'' (JPMXP1020200105),``Simulation for basic science: approaching the quantum era"  (JPMXP1020230411) and Joint Institute for Computational Fundamental Science (JICFuS). This work is supported in part by JSPS KAKENHI Grant Number 20H01907. I. K. acknowledges JSPS KAKENHI (JP20K0396). H.F. acknowledges JSPS Kakenhi grant numbers JP23K22490 and JP25K07283. T. K. acknowledges JSPS Kakenhi grant numbers 23K20846 and 25K01007. The simulations are performed on the supercomputer ``Fugaku'' at RIKEN Center for Computational Science (HPCI project hp240295,hp230207,hp200130, hp210165, hp220174 and Usability Research ra000001). J. G. and Y. Z. acknowledge support by the Deutsche Forschungsgemeinschaft (DFG, German Research Foundation) through the CRC-TR 211 ``Strong-interaction matter under extreme conditions''--project number 315477589--TRR 211. They also acknowledge support through DFG project number 460248186 (PUNCH4NFDI). J.G. would like to thank the long-term workshop on HHIQCD2024 at the Yukawa Institute for Theoretical Physics (YITP-T-24-02) for giving him a chance to deepen his ideas.

\section*{Data Availability}
All data from our calculations, presented in the figures of this paper, can be found in Ref.~\cite{Goswami:2026dataset}. We used the Grid~\cite{Boyle:2015tjk,githubGrid} for configuration generations and Bridge++~\cite{Ueda:2014rya,Aoyama:2023tyf, githubBridge} for measurements. We use AnalysisToolbox \cite{Altenkort:2023xxi,githubAnalysistoolbox} for statistical data analysis.

\appendix
\section{Trace identities and \(Z\)-combinations}
	The derivatives of $\ln \det D$ with respect to the dimensionless lattice chemical potential
	$\bar{\mu}$ are related to traces involving the inverse Dirac operator as
	\begin{align}
		\frac{\partial \ln \det D}{\partial \bar{\mu}} 
		&= \Tr \left( D^{-1} \frac{\partial D}{\partial \bar{\mu}} \right), \\
		\frac{\partial^2 \ln \det D}{\partial \bar{\mu}^2} 
		&= \Tr \left( D^{-1} \frac{\partial^2 D}{\partial \bar{\mu}^2} \right) 
		- \Tr \left( 
		D^{-1} \frac{\partial D}{\partial \bar{\mu}} 
		D^{-1} \frac{\partial D}{\partial \bar{\mu}} 
		\right), \\
		\frac{\partial^3 \ln \det D}{\partial \bar{\mu}^3} 
		&= \Tr \left( D^{-1} \frac{\partial^3 D}{\partial \bar{\mu}^3} \right) 
		- 3 \Tr \left( 
		D^{-1} \frac{\partial D}{\partial \bar{\mu}} 
		D^{-1} \frac{\partial^2 D}{\partial \bar{\mu}^2} 
		\right) \nonumber \\
		&\quad + 2 \Tr \left( 
		D^{-1} \frac{\partial D}{\partial \bar{\mu}} 
		D^{-1} \frac{\partial D}{\partial \bar{\mu}} 
		D^{-1} \frac{\partial D}{\partial \bar{\mu}} 
		\right), 
		\label{eq:dermu3} \\
		\frac{\partial^4 \ln \det D}{\partial \bar{\mu}^4} 
		&= \Tr \left( D^{-1} \frac{\partial^4 D}{\partial \bar{\mu}^4} \right) 
		- 4 \Tr \left( 
		D^{-1} \frac{\partial D}{\partial \bar{\mu}} 
		D^{-1} \frac{\partial^3 D}{\partial \bar{\mu}^3} 
		\right) \nonumber \\
		&\quad - 3 \Tr \left( 
		D^{-1} \frac{\partial^2 D}{\partial \bar{\mu}^2} 
		D^{-1} \frac{\partial^2 D}{\partial \bar{\mu}^2} 
		\right) \nonumber \\
		&\quad + 12 \Tr \left( 
		D^{-1} \frac{\partial D}{\partial \bar{\mu}} 
		D^{-1} \frac{\partial D}{\partial \bar{\mu}} 
		D^{-1} \frac{\partial^2 D}{\partial \bar{\mu}^2} 
		\right) \nonumber \\
		&\quad - 6 \Tr \left( 
		D^{-1} \frac{\partial D}{\partial \bar{\mu}} 
		D^{-1} \frac{\partial D}{\partial \bar{\mu}} 
		D^{-1} \frac{\partial D}{\partial \bar{\mu}} 
		D^{-1} \frac{\partial D}{\partial \bar{\mu}} 
		\right), 
		\label{eq:dermu4}
	\end{align}
Here, the \(Z\) terms are derived from combinations of \(D_n\) terms defined in Eq. (\ref{eq:Dn}):
\begin{align}
Z_2^f &= \langle (D_1^{f})^2 \rangle + \langle D_2^f \rangle, \\
Z_{11}^{fg} &= \langle D_1^f D_1^g \rangle, \\
Z_{22}^{fg} &= \langle (D_1^{f})^2 (D_1^{g})^2 \rangle + \langle (D_1^{f})^2 D_2^g \rangle + \langle (D_1^{g})^2 D_2^f \rangle + \langle D_2^f D_2^g \rangle, \\
Z_{31}^{fg} &= \langle (D_1^f)^3 D_1^g \rangle + 3\langle D_1^g D_1^f D_2^f\rangle + \langle D_1^g D_3^f\rangle, \\
Z_{211}^{fgh} &= \langle (D_1^{f})^2 D_1^g D_1^h \rangle + \langle D_2^f D_1^g D_1^h \rangle, \\
Z_4^f &= \langle (D_1^{f})^4 \rangle + 6 \langle (D_1^{f})^2 D_2^f \rangle + 4 \langle D_1^f D_3^f \rangle + 3 \langle (D_2^{f})^2 \rangle + \langle D_4^f \rangle.
\end{align}

\section{Blocking checks}
\label{app:resampling}

Measurements are separated by 100 molecular-dynamics trajectories. To test the stability of the jackknife errors, we group consecutive measurements into blocks corresponding to $b=100$, 200, and 400 trajectories. The figures show the jackknife error at block size $b$, normalized to that at the original measurement spacing, $\sigma_b/\sigma_{100}$.

\begin{figure}[htbp]
    \centering
    \includegraphics[width=0.64\textwidth]{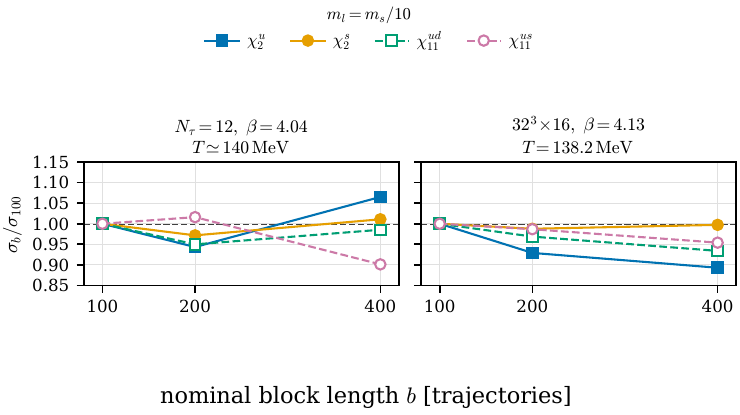}
    \caption{Blocking stability of the jackknife errors for $\chi_2^u$, $\chi_2^s$, $\chi_{11}^{ud}$, and $\chi_{11}^{us}$ on the two $m_l=m_s/10$ ensembles relevant near $T\simeq140~\mathrm{MeV}$: $N_\tau=12$, $\beta=4.04$, $T\simeq140~\mathrm{MeV}$, and $N_\tau=16$, $\beta=4.13$, $T=138.2~\mathrm{MeV}$. The plotted quantity is $\sigma_b/\sigma_{100}$ for block sizes $b=100$, 200, and 400 molecular-dynamics trajectories.}
    \label{fig:appendix_resampling_second_order}
\end{figure}

The blocking results for the second-order observables are shown in Fig.~\ref{fig:appendix_resampling_second_order}. For these observables, the errors vary at roughly the 10\% level and, importantly, show no systematic growth as the block size increases. This blocking test therefore gives no indication that the statistical errors near $T\simeq140~\mathrm{MeV}$ are substantially underestimated because of residual autocorrelations; it does not establish that autocorrelations are absent. For the diagonal susceptibilities on these two ensembles, configuration-level bootstrap estimates agree with the ordinary jackknife errors within approximately 3\%, providing an additional check using an independent resampling prescription.

\begin{figure}[htbp]
    \centering
    \includegraphics[width=0.44\textwidth]{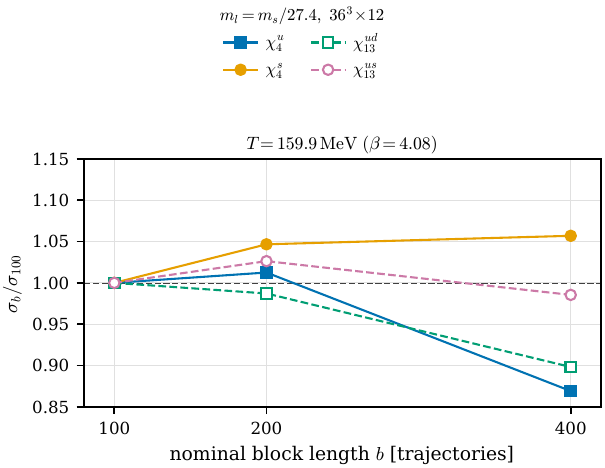}
    \caption{Representative blocking stability of the jackknife errors for selected fourth-order quark-number susceptibilities on the physical-mass ensemble, $m_l=m_s/27.4$, with volume $36^3\times12$, $\beta=4.08$, and $T=159.9~\mathrm{MeV}$. The plotted quantity is $\sigma_b/\sigma_{100}$ for block sizes $b=100$, 200, and 400 molecular-dynamics trajectories.}
    \label{fig:appendix_resampling_fourth_order}
\end{figure}

The corresponding fourth-order blocking results are shown in Fig.~\ref{fig:appendix_resampling_fourth_order}. For this representative fourth-order check, the errors remain stable at approximately the 10--15\% level as the block size changes from 100 to 400 trajectories, with no systematic increase common to the different observables.
\clearpage
\bibliographystyle{JHEP}
\bibliography{bibliography}

\end{document}